\documentclass[
 reprint,
 amsmath,amssymb,
 aps,
]{revtex4-2}

\pdfoutput=1

\usepackage{graphicx}
\usepackage{dcolumn}
\usepackage{bm}
\usepackage{float}
\usepackage[utf8]{inputenc}
\usepackage{booktabs}
\usepackage{tabularx}
\usepackage{colortbl}
\renewcommand{\vec}[1]{\mathbf{#1}}
\usepackage{etoolbox}
\usepackage{titlesec}
\usepackage{braket}
\usepackage{xcolor}
\usepackage[font=small]{caption}

\titlespacing*{\section}{0pt}{0.5\baselineskip}{0.5\baselineskip}
\titlespacing*{\subsection}{0pt}{0.5\baselineskip}{0.4\baselineskip}

\makeatletter
\def\maketitle{
\@author@finish
\title@column\titleblock@produce
\suppressfloats[t]}
\makeatother

\begin{document}
\title{Erratum: UPconversion Loop Oscillator Axion Detection experiment: A precision frequency interferometric axion dark matter search with a Cylindrical Microwave Cavity}

\author{Catriona A. Thomson}
\email{catriona.thomson@research.uwa.edu.au}
\affiliation{ARC Centre of Excellence for Engineered Quantum Systems and ARC Centre of Excellence for Dark Matter Particle Physics, Department of Physics, University of Western Australia, 35 Stirling Highway, Crawley, WA 6009, Australia.}
\author{Ben T. McAllister}
\affiliation{ARC Centre of Excellence for Engineered Quantum Systems and ARC Centre of Excellence for Dark Matter Particle Physics, Department of Physics, University of Western Australia, 35 Stirling Highway, Crawley, WA 6009, Australia.}
\author{Maxim Goryachev}
\affiliation{ARC Centre of Excellence for Engineered Quantum Systems and ARC Centre of Excellence for Dark Matter Particle Physics, Department of Physics, University of Western Australia, 35 Stirling Highway, Crawley, WA 6009, Australia.}
\author{Eugene N. Ivanov}
\affiliation{ARC Centre of Excellence for Engineered Quantum Systems and ARC Centre of Excellence for Dark Matter Particle Physics, Department of Physics, University of Western Australia, 35 Stirling Highway, Crawley, WA 6009, Australia.}
\author{Michael E. Tobar}
\email{michael.tobar@uwa.edu.au}
\affiliation{ARC Centre of Excellence for Engineered Quantum Systems and ARC Centre of Excellence for Dark Matter Particle Physics, Department of Physics, University of Western Australia, 35 Stirling Highway, Crawley, WA 6009, Australia.}
\date{\today}
\maketitle
\onecolumngrid
\vspace{-7mm}
We have found a sign error in our sensitivity calculation. Our results still experimentally demonstrate the feasibility of exploiting upconversion/downconversion to search for axions, but the sensitivity to low mass axions claimed in~\cite{Thomson2019,Goryachev2019}, and quoted in~\cite{PhysRevD.103.015034,billard2021direct,hall2021predictions}, must be substantially reduced by four orders of magnitude to $3\times 10^{-3}$ 1/GeV between 7.44 - 19.38 neV and by up to $11$ orders of magnitude for the lowest axion masses in the theoretical projections in the left panel of Fig.~3 in the main text. Conversely, the sensitivity to high mass axions must be increased to $10^{-2}$ 1/GeV between 74.4 - 74.5 $\mu$eV. With this correction, our results show that calculating sensitivity using the photon-axion interaction Hamiltonian is consistent with using a derivative coupling to the axion, showing suppressed sensitivity as $m_a \to 0$. This is consistent with other work on axion searches in excited cavities~\cite{berlin2020axion,PhysRevD.103.075007}. Thus, our conclusion that measuring axion induced frequency shifts was much better than measuring axion induced power, was in fact wrong.

\begin{figure}[h]
\centering
\includegraphics[width=11.8cm]{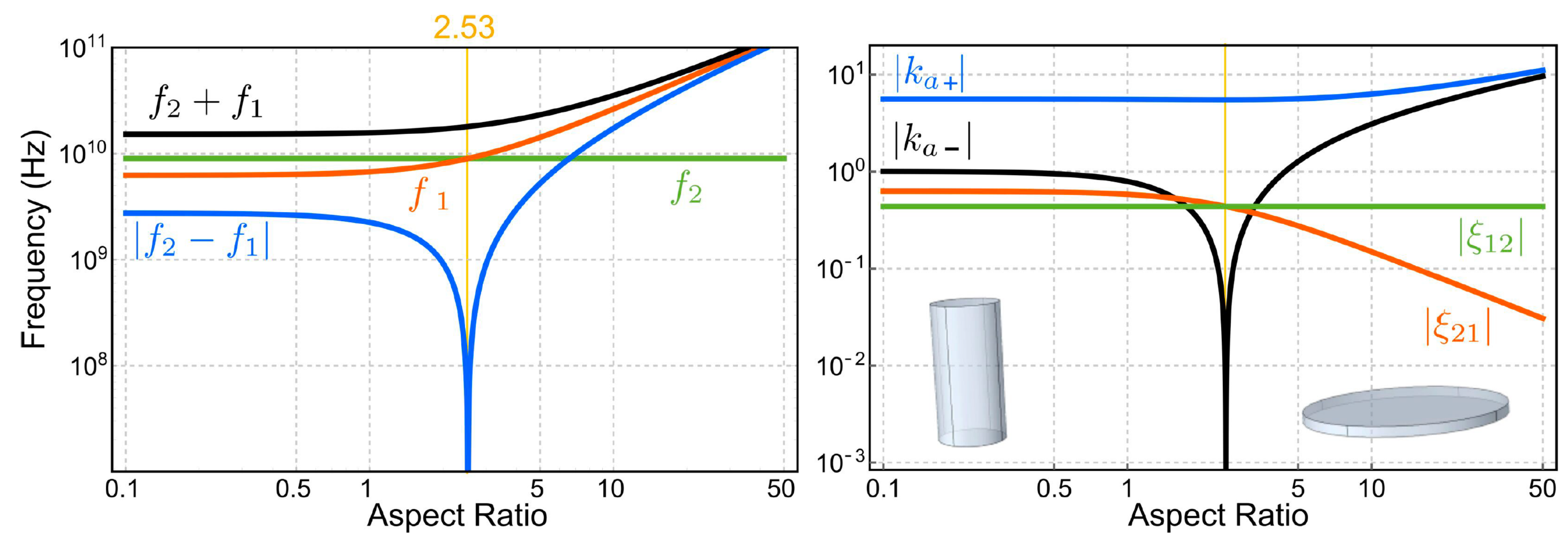}
\caption[XiGraph]{Corrected Fig.~3 in the supplementary material of $k_{a\pm}$ conversion factors versus Aspect Ratio. Upconversion sensitivity is suppressed in the low axion mass case when the Aspect Ratio is 2.53, as in the reported experiment.}
\label{XiGraph}
\end{figure}

The sign error occurs in Eq.~(11) of the Supplemental Material, where the normalized unit vectors for the $\text{TE}_{\text{0,1,1}}$ mode are given. Here, the negative sign before the $z$-component of $\mathbf{b}_1$ was missed. The correct expressions are:
\begin{align}
\mathbf{e}_{1} = -\frac{\sqrt{2}J_1\left(\frac{\chi_{01}'}{a}r\right) \sin \left(\frac{\pi  z}{L}\right) }{J_0(\chi_{01}')} \ \hat{\varphi}, ~~ \text{and} ~~
\mathbf{b}_{1}=\frac{\sqrt{2}\frac{a\pi}{L\chi_{01}'}J_1\left(\frac{ \chi_{01}'}{a}r\right)\cos \left(\frac{\pi  z}{L}\right) }{\sqrt{1+\left(\frac{a\pi}{L\chi_{01}'}\right)^2}J_0(\chi_{01}')} \ \hat{r} - 
\frac{\sqrt{2}J_0\left(\frac{ \chi_{01}'}{a}r\right)\sin \left(\frac{\pi  z}{L}\right) }{\sqrt{1+\left(\frac{a\pi}{L\chi_{01}'}\right)^2}J_0(\chi_{01}')} \ \hat{z}.
\end{align}
These expressions are used to derive the form factors $\xi_{\pm}$ and hence the conversion factors $k_{a\pm}$, which determine the sensitivities to downconverted (sum frequency) and upconverted (difference frequency) axions respectively. The error therefore exchanges the values of $\xi_\pm$, which are correctly written as,
\begin{equation}
        \begin{split}
\xi_{-}=\xi_{12}-\xi_{21}=\xi_{12}\left(1-\frac{f_2}{f_1}\right),~\text{and}~\xi_{+}=-(\xi_{12}+\xi_{21})=-\xi_{12} \left(1+\frac{f_2}{f_1}\right),  \\
        	\end{split}
	\label{pmoverlap}
\end{equation}
which is the corrected Eq. (14) in the Supplemental Material. Also, the error exchanges the values of $k_{a\pm}$,
\begin{equation}
k_{a\pm}^2=\frac{32 \chi_{01}'^2}{\left(\chi_{02}^2-\chi_{01}'^2\right)^2}\frac{\beta_{1} P_{1} Q_{L1} (\beta_2+1)^2}{\beta_{2} P_{2} Q_{L2}(\beta_1+1)^2}\frac{(f_2\pm f_1)^2}{f_1f_2},
\label{sens2}
\end{equation}
which is the corrected Eq. (18) in the Supplemental Material. The quoted conversion factors in Tab.~1 of the main text should be reversed, with $k_{a+}=5.5$ and $k_{a-}=8.4\times 10^{-4}-1.1\times 10^{-3}$, and Fig. 3 in the supplementary material should be replaced by Fig.\ref{XiGraph}.

Hence, the quoted SNRs for the free running loop oscillator (Eq.~(8) and in the main text and Eq.~(22) in the Supplemental Material) and the stabilized loop oscillator (Eq.~(25) in the Supplemental Material) are written incorrectly. When the modes are closely tuned, the equations give the sensitivity to downconverted axions, with $f_{a_{-}}$ replaced with $f_{a_{+}}$. Thus, given the condition that $f_1=f_2+\delta f_{12}$ is tuned slightly away from $f_2$, so that $\delta f_{12} / f_2 \ll 1$, and assuming $\beta_1 = 1$, the corrected Eq.~(8) in the main text (or Eq.~(22) in the Supplemental Material) is
\begin{figure}[t]
\centering
\includegraphics[width=17cm]{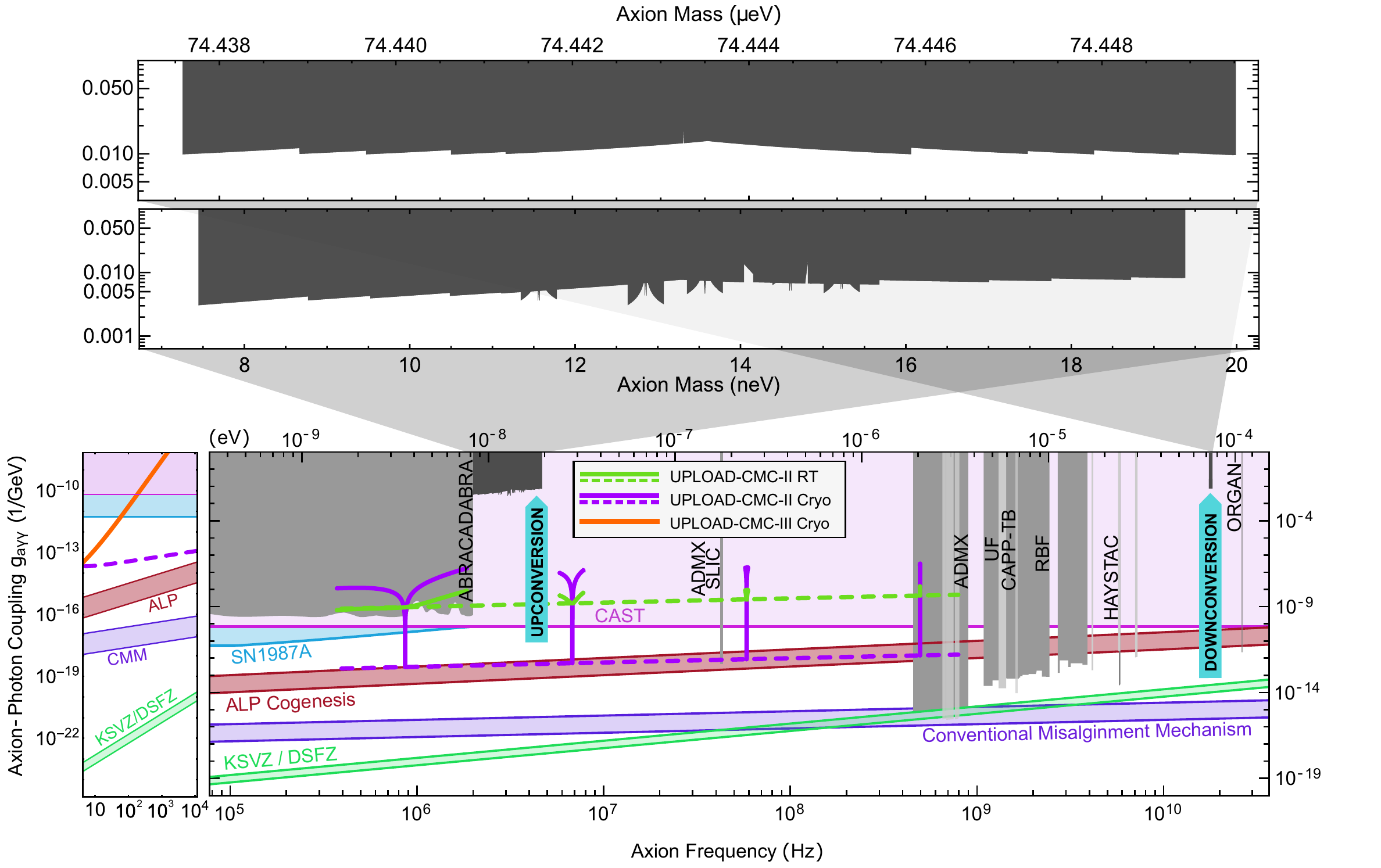}
\caption[ExclusionPlot]{Updated 95$\%$ confidence exclusion zones for $g_{a\gamma\gamma}$ for the measured upconversion and downconversion mass ranges (above), with CAST's helioscope limits in pink and QCD axion models in green (KSVZ and DSFZ) \cite{CAST2007,Graham2016}, purple (conventional ALP misalignment), and red (ALP cogenesis) \cite{Co2020,PhysRevLett.124.251802}.  Corrected upconversion limits are dashed, assuming 1 Hz tuning steps (30 days per Hz): UPLOAD-CMC-II-RT, a copper resonator with frequency stabilized (FS) loops at room temperature (RT); and UPLOAD-CMC-II-Cryo, cryogenic Nb with FS loops. We also present examples of 30 day measurements covering 1 MHz in Fourier space, in bold. The left sub-plot displays (in orange) a 30 day measurement projection for UPLOAD-CMC-III-Cryo, which is simply the cryogenic FS setup when modes are tuned just 5 Hz apart. The results are compared to ADMX \cite{Du2018,Boutan2019,Braine2020,Bartram2020}, ORGAN \cite{McAllister2017}, ABRACADABRA \cite{Henning2019}, ADMX-SLIC \cite{ADMXSLIC}, HAYSTAC \cite{Brubaker2017}, UF \cite{Hagmann1990}, CAPP-8TB \cite{Kultask20}, and RBF \cite{Wuensch1989}.}
\label{ExclusionPlot}
\end{figure}
\begin{equation}
SNR_{-}=g_{a\gamma\gamma}\frac{2.7\left(\frac{10^6t}{f_{a_{-}}}\right)^\frac{1}{4}\sqrt{\rho_{DM}c^3}}{2\pi f_{a_{-}}} \sqrt{\frac{Q_{L2}P_{amp} (\beta_2+1)^2}{(Fk_BT_0)\beta_{2} P_{2}}}\sqrt{P_{1} Q_{L1}}\sqrt{\frac{1}{\left(2Q_{L2}\frac{f}{f_2}\right)^2+1}}\left(\frac{|\delta f_{12}|}{\sqrt{2}f_2}\right),
\label{eq:FreeSNR}
\end{equation}
which has an extra term at the end proportional to the detuning between the two modes. Likewise Eq. (25) becomes,
\begin{equation}
SNR_{-}=g_{a\gamma\gamma}\frac{3.9\left(\frac{10^6t}{f_{a_{-}}}\right)^\frac{1}{4}\sqrt{\rho_{DM}c^3}}{2\pi f_{a_{-}}}\sqrt{\frac{Q_{L1}P_{1}}{k_bT_{RS}}}\sqrt{\frac{\beta_2Q_{L2}}{\left(2Q_{L2}\frac{f}{f_2}\right)^2+1}}\left(\frac{|\delta f_{12}|}{\sqrt{2}f_2}\right).
\end{equation}
Thus, we present the corrected exclusion plot for $g_{a\gamma\gamma}$ according to the corrected SNR equations in Fig.~\ref{ExclusionPlot}.

We are very grateful to Kevin Zhou for meticulously going through our paper and finding our mistake, and reading this correction making sure of its validity. This work was funded by the ARC Centre of Excellence for Engineered Quantum Systems, CE170100009, and Dark Matter Particle Physics, CE200100008.


\preprint{APS/123-QED}

\clearpage
\counterwithin{figure}{section}

\renewcommand\thefigure{\arabic{figure}}
\setcounter{figure}{0}

\title{\textit{Corrected:} UPconversion Loop Oscillator Axion Detection experiment: A precision frequency interferometric axion dark matter search with a Cylindrical Microwave Cavity}

\author{Catriona A. Thomson}
 \email{catriona.thomson@research.uwa.edu.au}
 \author{Ben T. McAllister}
\author{Maxim Goryachev}
\author{Eugene N. Ivanov}
\author{Michael E. Tobar}
 \email{michael.tobar@uwa.edu.au}

 \affiliation{ARC Centre of Excellence for Engineered Quantum Systems and ARC Centre of Excellence for Dark Matter Particle Physics, Department of Physics, University of Western Australia, 35 Stirling Highway, Crawley, WA 6009, Australia.}

\date{\today}

\begin{abstract}
First experimental results from a room-temperature table-top phase-sensitive axion haloscope experiment are presented. The technique exploits the axion-photon coupling between two photonic resonator-oscillators excited in a single cavity, allowing low-mass axions to be upconverted to microwave frequencies, acting as a source of frequency modulation on the microwave carriers. This new pathway to axion detection has certain advantages over the traditional haloscope method, particularly in targeting axions below 1 $\mu$eV (240 MHz) in energy where high volume magnets are necessary. At the heart of the dual-mode oscillator, a tunable cylindrical microwave cavity supports a pair of orthogonally polarized modes ($\text{TM}_{\text{0,2,0}}$ and $\text{TE}_{\text{0,1,1}}$), which, in general, enables simultaneous sensitivity to axions with masses corresponding to the sum and difference of the microwave frequencies. The results place axion exclusion limits between 7.44 - 19.38 neV, excluding a minimal coupling strength above $3\times 10^{-3}$ 1/GeV, and between 74.4 - 74.5 $\mu$eV, excluding a minimal coupling strength above $10^{-2}$ 1/GeV, after a measurement period of two and a half hours. We show that a state-of-the-art frequency-stabilized cryogenic implementation of this technique may achieve competitive limits in a large range of axion-space.
\end{abstract}

\maketitle

The nature of dark matter in our universe has long been a looming question in physics and the focus of extensive experimental efforts today. Weakly-interacting sub-eV particles (WISPs) are becoming increasingly suspect in the wake of sustained non-detection by high-mass experiments \cite{Graham2016}. The axion, a theorized Nambu-Goldstone boson emerging from the Peccei-Quinn (PQ) solution to the strong charge-parity (CP) problem in quantum chromodynamics (QCD) \cite{Peccei1977,Bertone2010,Wilczek1978}, is a popular candidate for cold dark matter, with a mass poorly constrained by theory; several orders of magnitude are available for exploration \cite{Turner1990,Peccei2006,Ballesteros2016,Jaeckel2007}. The majority of axion experiments that aim to detect the QCD axion are `haloscopes', which are sensitive to power deposition from the conversion of galactic halo axions into photons through the inverse Primakoff effect, as predicted by the axion-augmented QCD Lagrangian \cite{Sikivie1984,Younggeun18,Tobar2019,TOBAR2020}. Experiments such as the well-established Axion Dark Matter eXperiment (ADMX) \cite{Du2018,Boutan2019,Braine2020}, the ORGAN Experiment \cite{McAllister2017}, HAYSTAC \cite{Droster2019} and CULTASK \cite{Chung2016} have hitherto depended upon low-noise microwave receivers and cryogenic cooling to detect excess real microwave photons produced in a low-loss microwave cavity surrounded by a strong DC magnetic field.

\begin{figure}[t]
\centering
\includegraphics[width=7cm]{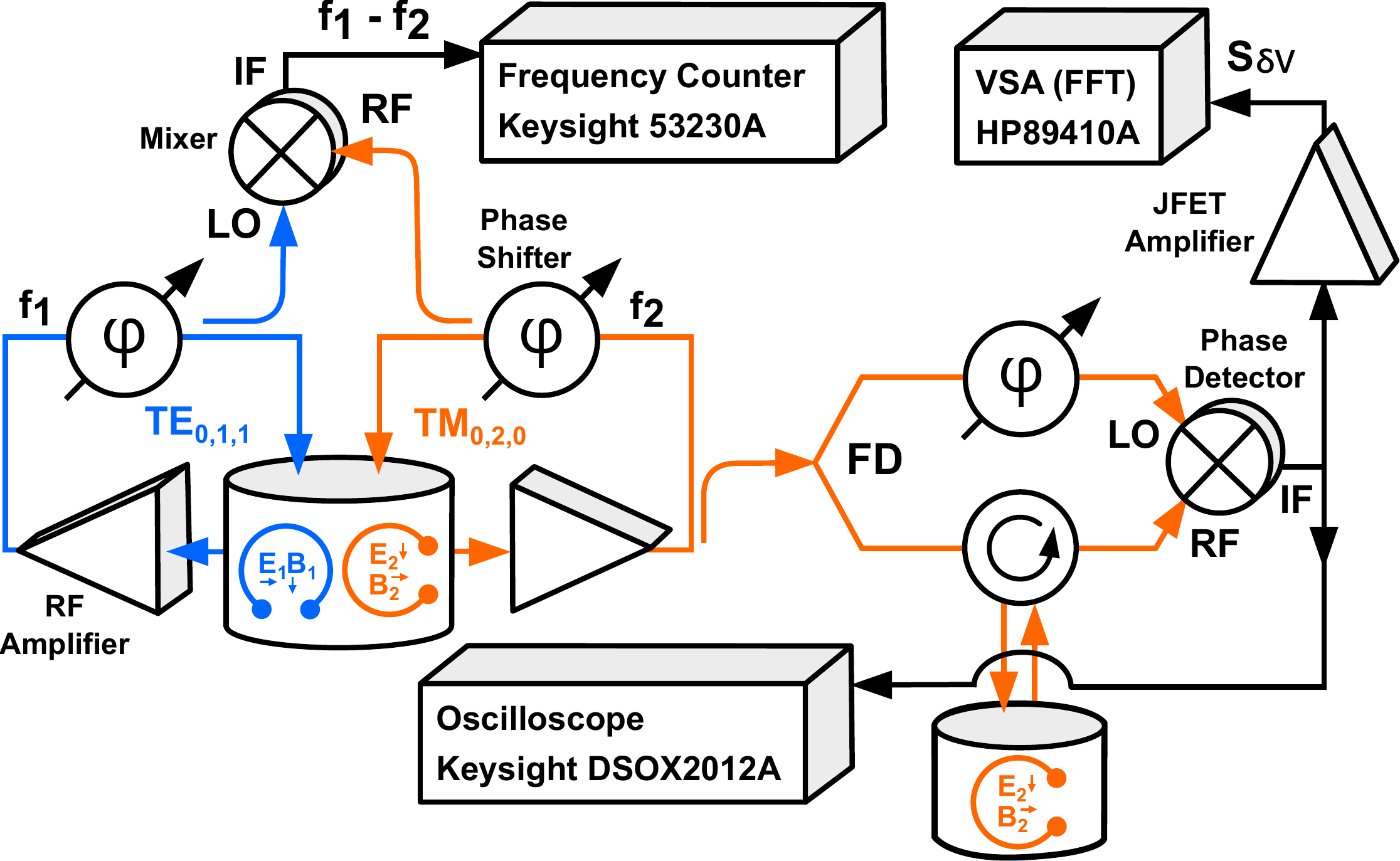}
\caption[UpDownConversion]{UPLOAD-CMC schematic with frequency noise readout system. Loop oscillators support the microwave modes in the main resonator allowing axion conversion while a frequency counter records $|f_2-f_1|$. The TM mode acts as the readout mode with constant $f_2$. The frequency noise is continuously scanned with a high-Q frequency discriminator (FD) based on a a duplicate cavity, creating the frequency-phase dispersion required for voltage to frequency noise conversion of the mixer output.}
\label{Schematic}
\end{figure}

Our approach similiarly employs an electromagnetic resonance in a low-loss microwave cavity, but instead of inducing axion-photon conversion by applying an external DC magnetic field, we excite another resonance within the cavity to act as the second source of photons. The spatial overlap between the E-field of one mode in the cavity with the B-field of the other induces the desired coupling. We dub this configuration the AC haloscope, in reference to the DC field normally employed in haloscope experiments. The configuration was conceived from the observation that while most haloscopes attempt to scatter axions off a virtual photon source, the Primakoff process also generates products in the presence of real photons, as noted by Sikivie in 2010 \cite{Sikivie2010}. An axion with an energy corresponding to the sum or difference frequency of the photons is expected to interact and may be detected via frequency \cite{Goryachev2019} or power measurements \cite{Sikivie2010,Lasenby2020,Lasenby2020b}. Uniquely, the AC haloscope allows one to search for an axion signal imprinted in the phases of the photon modes, placing it in a new class of haloscopes, focused on frequency metrology, instead of power detection \cite{Goryachev2019}. Frequency techniques have been used in the past for some of the best tests of fundamental physics, including variations in fundamental constants and local position invariance \cite{Turneaure1983,Tobar2010,Tobar2013}, as well as tests of Lorentz invariance violation \cite{Nagel2015,Lo2016,Goryachev2018}, with proven long-term performance of up to eight years \cite{Tobar2013}, and if designed properly the sensitivity will be determined by the white frequency noise floor of the frequency stabilization system \cite{Nagel2015}.

The experiment in this work comprises a resonant cavity supporting two spatially overlapping microwave modes via two free-running loop oscillators. Our chosen geometry has enhanced sensitivity to downconversion (photon sum) and less sensitivity to upconversion (photon difference), which is detailed in supplemental material. As a result, we present the UPLOAD (UPconversion Loop Oscillator Axion Detection) experiment using a Cylindrical Microwave Cavity (UPLOAD-CMC), with the schematic shown in Fig.~\ref{Schematic}. By searching for frequency deviations of the carrier frequencies of the oscillators, this experiment can be configured to cover a large portion of unexplored low-mass axion-space, below the ADMX mass range, in principle from DC to 240 MHz ($<1 \mu$eV).

Experimentally, this design enables a coherent phase or frequency modulation induced by the axion-photon interaction to be scanned. The large signal effective noise temperature of the sustaining loop amplifier limits the system's frequency stability and therefore our ability to sense the predicted modulation. The results produced in this experiment place limits on axion-photon coupling in the MHz and GHz rangesr by observing the absence of axion-induced frequency modulation in orthogonally oriented photonic modes oscillating within a small cylindrical copper cavity at room temperature.

The axion-photon interaction Hamiltonian density is familiarly parameterized as
\begin{equation} \label{eq1}
 \mathcal{H}_{int} = \epsilon_{0}{c}{g}_{a\gamma\gamma}a\vec{E}\cdot\vec{B},
\end{equation}
where $\vec{E}$ and $\vec{B}$ represent the electric and magnetic fields, $a$ is the psuedoscalar axion-like field, $c$ is the speed of light, $\epsilon_{0}$ is the permittivity of free space, and ${g}_{a\gamma\gamma}$ is the axion-photon coupling strength \cite{Ringwald2012}. The Hamiltonian can be expanded to account for the electric and magnetic field distributions of two photonic modes. The interaction may be rewritten in terms of creation and annihilation operators as
    \begin{equation}
    \begin{split}
        {\text{H}}_{int} ={i}\pi\hbar{g}_{a\gamma\gamma}a\sqrt{f_{1}f_{2}}\big[\xi_{-}({c_1}&{c_2^{\dagger}}-{c_1^{\dagger}}{c_2}) \\
        & +\xi_{+}({c_1^{\dagger}}{c_2^{\dagger}}-{c_1}{c_2}) \big],
         \label{eq2}
         \end{split}
    \end{equation}
where $\xi_{\pm} =-(\xi_{21}\pm\xi_{12})$ encodes the necessary mutual overlapping of the electric and magnetic field components of the two modes (see \cite{Goryachev2019} as well as the supplemental material for more details) and $f_1$ and $f_2$ are the resonant frequencies of the photon modes. We find that the following regimes satisfy the Hamiltonian
    \begin{equation} 
        \begin{split}
       f_{a} ={} & f_{2} + f_{1} \,\,\, \text{ axion downconversion}\\
       f_{a} ={} &\lvert f_{2}\, {-}\, f_{1}\rvert \,\,\, \text{ axion upconversion}
        \end{split}
        \label{eq3}
    \end{equation}
where $f_{1,2}\ne 0$. In essence, the axion field couples two microwave oscillators together when the axion frequency is at the sum or difference of the microwave frequencies, creating frequency or phase shifts in the microwave frequencies, arising from this trilinear coupling. 

The phase shifts arise when we relax the frequency relations from Eqn. \ref{eq3} to
    \begin{equation} \label{eq4}
f_{a} = f_{2}+f_{1} \pm f \ \ \text{or}  \ \  f_{a} =\lvert f_{2}-f_{1}\rvert \pm f,
    \end{equation}
where, $f \ll f_{1,2}$ represents a small offset. In this new relation, the axion amplitude becomes a slowly varying parameter in the equations of motion, inducing phase noise in the Fourier spectrum of both cavity modes, located at the offset frequency. A derivation of the transfer function from this slowly varying amplitude to resultant phase modulation can be found in Appendix B of \cite{Goryachev2019}. Note, in the supplemental material we derive the resulting signal in terms of a fractional \textit{frequency} modulation (related to phase noise), and use this approach throughout this work.
\begin{figure*}[t]
\centering
\includegraphics[width=16.5cm]{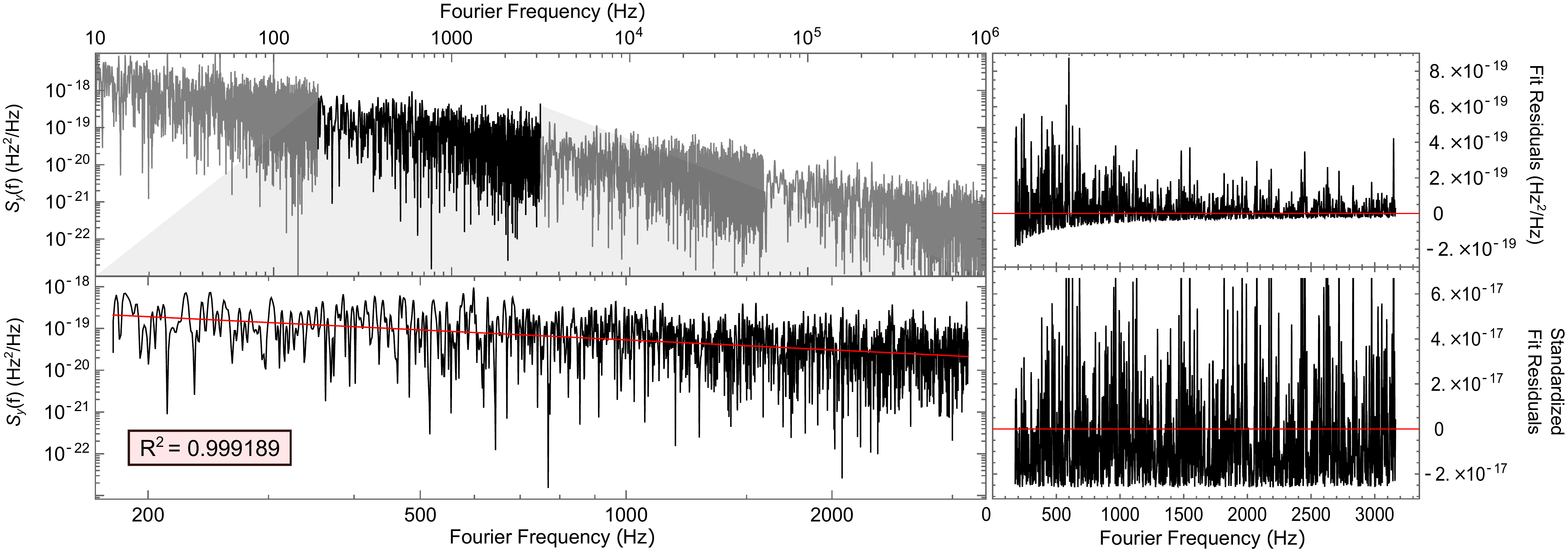}
\caption[UpDownConversion]{ \textbf{Top-left:} Filtered spectral density of fractional frequency noise at Fourier frequencies about the normal mode frequency, $f_{\text{TM}}$, for $f_{\text{TM}}$ = 8.9989 GHz and $f_{\text{TE}}$ = 9.00241 GHz. \textbf{Bottom-left:} The second decade of data, with inverse power law fit. Decades were analyzed independently. \textbf{Top-right:} Residuals from the spectral density of fractional frequency noise fit, examinable as axion-induced excesses.  \textbf{Bottom-right:} Standardized residuals (henceforth referred to as ``search spectrum amplitudes''), ready to be interrogated with the axion-search procedure.}
\label{NoiseFit}
\end{figure*}

This induced modulation appears in the spectral density of fractional frequency fluctuations of the readout oscillator in the form
\begin{equation}
    \begin{split}
    S_{Ay_{2}}(f)_{\pm} = g_{a\gamma\gamma}^2k_{a\pm}^2S_A(f), \ \  k_{a\pm}^2= \pi^2\frac{f_1P_{c1}}{f_2P_{c2}}\xi_{\pm}^2, 
  \end{split}
    \label{SySAa}
\end{equation}
where $P_{c1}/P_{c2}$ is the ratio of pump mode (subscript 1) and readout mode (subscript 2) circulating power in the resonator (see Fig. \ref{Schematic}) and $\xi_{\pm}$ is the geometric overlap factor, describing the efficiency of coupling between the two electromagnetic modes. Here $k_{a\pm}$ is the conversion ratio from axion theta angle, $\theta=g_{a\gamma\gamma} a$, to fractional frequency deviation, $y=\frac{\delta f}{f_0}$, with calculated values for this run shown in Tab.~\ref{tab:params}. Here, the axion (or axion-like particle) field may be considered as a spectral density of narrow-band noise, centred at a frequency equivalent to the axion mass and broadened due to cold dark matter virilization to give a linewidth of ${10^{-6}{f_{a}}}$, and is denoted as ${S_{A}(f)}$ (kg/s/Hz) \cite{Lentz2017}. This must compete against the oscillator noise given by Leeson's model \cite{Leeson1966,Lance1984}, which is typically of the form (see the supplemental material for more details and how to calculate SNR) 
\begin{equation}
S_{\phi_{2}}(f)_{osc}= S_{\phi_{2}} (f)_{amp}\left(1+\left(\frac{\Delta f_{L_2}}{2{f}}\right)^2\right),
     \label{eqS1}
\end{equation}
which can be converted to fractional frequency fluctuations via
\begin{equation}
S_{y_2}(f)_{osc} =\left(\frac{f}{f_2}\right)^2S_{\phi_{2}}(f)_{osc}{ .}
     \label{eqS1}
\end{equation}

Here, $S_{\phi_2}(f)_{amp}$ is the phase noise of the amplifier in the feedback loop of the readout oscillator and $\Delta f_{L_2}$ is the readout mode full bandwidth. For our free-running experiment, we find the SNR to be
\begin{equation}
\begin{split}
SNR_{-}=&g_{a\gamma\gamma}\frac{2.7\left(\frac{10^6t}{f_{a_{-}}}\right)^\frac{1}{4}\sqrt{\rho_{DM}c^3}}{2\pi f_{a_{-}}} \sqrt{\frac{Q_{L2}P_{amp} (\beta_2+1)^2}{(Fk_BT_0)\beta_{2} P_{2}}}\\
&\sqrt{P_{1} Q_{L1}}\sqrt{\frac{1}{\left(2Q_{L2}\frac{f}{f_2}\right)^2+1}}\left(\frac{|\delta f_{12}|}{\sqrt{2}f_2}\right),
\end{split}
\label{eq:FreeSNR}
\end{equation}
explained in more detail in the supplemental material.

\begin{table}\centering
\caption{Experimental Microwave Oscillator Parameters} 
\begin{tabular*}{\linewidth}{l@{\extracolsep{\fill}}lll}
\toprule
        & $\text{TE}_{\text{0,1,1}}$ \text{(Mode 1 Fig.~\ref{Schematic})}   & $\text{TM}_{\text{0,2,0}}$ (Mode 2 Fig.~\ref{Schematic})  \\ \midrule\midrule
$Q_L$ 							 &    6000      &   4200       \\ 
$\beta_{\text{in}}$  		&     0.9        &      0.95        \\ 
$P_{\text{inc}}~~~~$ 		&   10 dBm   &   6 dBm     &                \\ 
$P_{\text{c}}$ 		&   48 dBm   &   42 dBm     &                \\ 
$f_0$ 						& 9.00168 - 9.00256 GHz  &   8.9988765 GHz  \\ 
$k_{a-}$  &	 $8.4\times 10^{-4}-1.1\times 10^{-3}$ &	 $8.4\times 10^{-4}-1.1\times 10^{-3}$ \\ 
$k_{a+}$ &	$5.5$ &	$5.5$  \\    \bottomrule
\end{tabular*}
\label{tab:params}
\end{table}

A cylindrical copper resonator was designed, in which two orthogonal modes could simultaneously oscillate. It was noted that a $\text{TM}_{0,2,0}$ mode offered a higher quality factor than a $\text{TM}_{0,1,0}$ mode without inducing significant mode crowding. A $\text{TE}_{0,1,1}$ mode was chosen for its wide frequency tuning range within the cavity and its optimal modal overlap with the chosen $\text{TM}_{0,2,0}$ mode. Specific parameters of the electromagnetic resonances are included in Tab.~\ref{tab:params} and details of the overlap function are given in the supplemental material.

Two free-running loop oscillators were constructed from the TE  and TM mode resonances, with the beat frequency under constant measurement via an RF mixer connected to a Keysight Frequency Counter (53230A). We chose to measure the frequency noise of the TM mode due to its relatively stationary frequency, which required minimal tuning of the frequency noise discrimination circuit, comprising a standard phase bridge, including a phase shifter, a double balanced mixer and a duplicate resonator included as the dispersive element. The reflected and phase-shifted signals were mixed in quadrature, such that the mixer output voltage was proportional to the frequency noise of the loop oscillator, including the frequency noise of the resonator where the axion signal is imprinted \cite{Shoaf1973}. Essentially, the spectral density of fractional frequency fluctuations of the loop oscillator can be inferred from the measured spectral density of voltage fluctuations output by the mixer, $S_v(f)$, by
\begin{equation}
S_{y_2}(f)_{osc}= \frac{S_{v_2}(f)}{f_2^2k_f^2G},
\label{SySv}
\end{equation}
where $f_2$ is the TM loop oscillator frequency and $k_f$ and $G$, conversion efficiency and gain of the post-amplifier. As an axion signal (Eqn.~\ref{SySAa}) must compete with the spread of this background to appear in the total frequency noise at the readout, frequency fluctuations in the oscillator and readout system are the limiting factor of this experiment.
    
\begin{figure*} 
    \includegraphics[width=16cm]{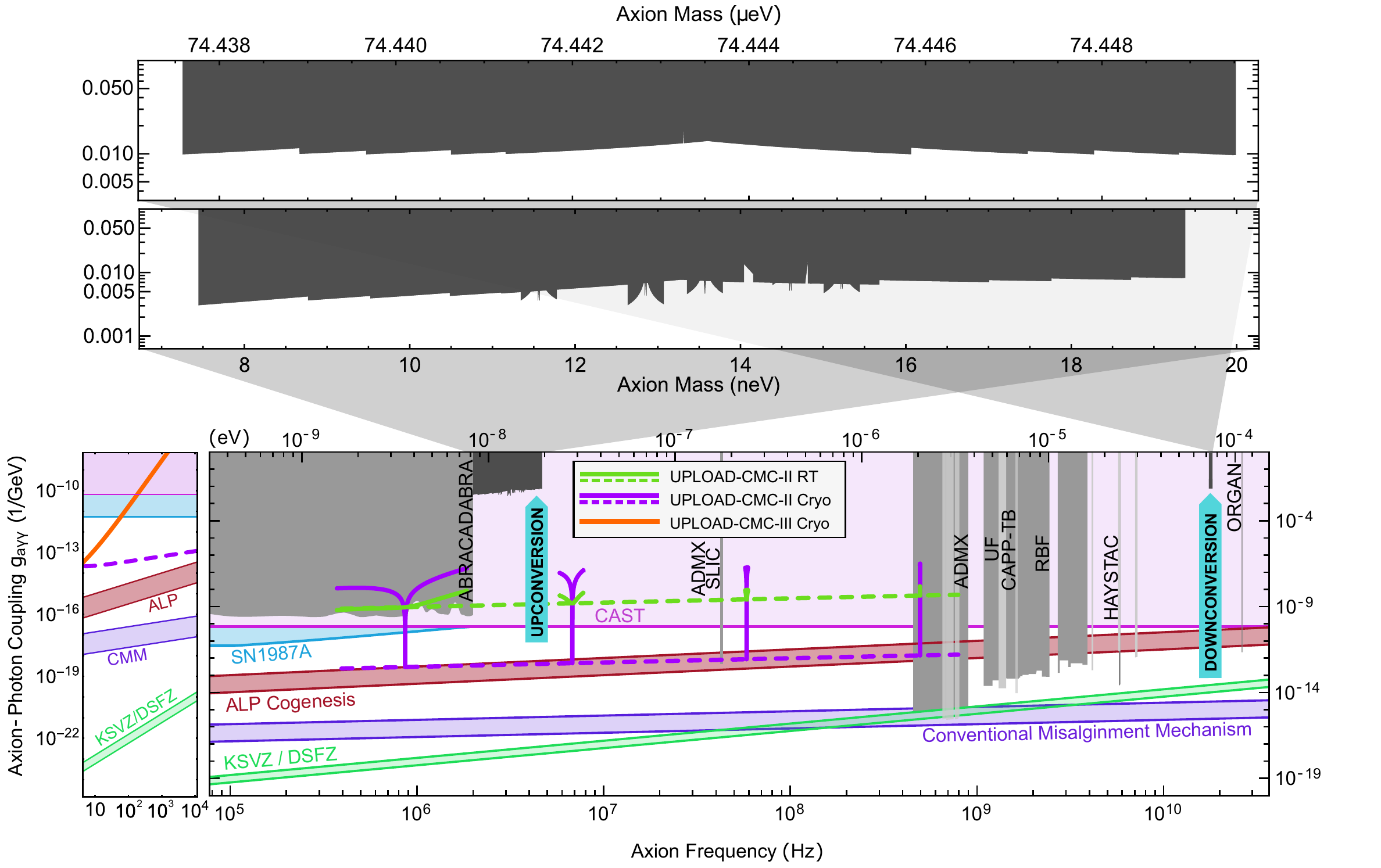}
\caption[ExclusionPlot]{95$\%$ confidence exclusion zones for $g_{a\gamma\gamma}$ (in natural units) for the measured upconversion and downconversion mass ranges (above), with CAST's helioscope limits in pink and QCD axion models in green (KSVZ and DSFZ) \cite{CAST2007,Graham2016}, purple (conventional ALP misalignment), and red (ALP cogenesis) \cite{Co2020,PhysRevLett.124.251802}.  Upconversion limits are dashed, assuming 1 Hz tuning steps (30 days per Hz): UPLOAD-CMC-II-RT, a copper resonator with frequency stabilized (FS) loops at room temperature (RT); and UPLOAD-CMC-II-Cryo, cryogenic Nb with FS loops. We also present examples of 30 day measurements covering 1 MHz in Fourier space, in bold. The left sub-plot displays (in orange) a 30 day measurement projection for UPLOAD-CMC-III-Cryo, which is simply the cryogenic FS setup when modes are tuned just 5 Hz apart. The results are compared to ADMX \cite{Du2018,Boutan2019,Braine2020,Bartram2020}, ORGAN \cite{McAllister2017}, ABRACADABRA \cite{Henning2019}, ADMX-SLIC \cite{ADMXSLIC}, HAYSTAC \cite{Brubaker2017}, UF \cite{Hagmann1990}, CAPP-8TB \cite{Kultask20}, and RBF \cite{Wuensch1989}.}
 \label{ExclusionPlot}
\end{figure*} 

Repeating measurements at slightly different detunings will cause a putative axion-induced signal to be correspondingly translated in the Fourier space of the measured noise spectrum. Therefore, the axion search involves combining several noise spectra and looking for evidence of a signal consistently satisfying Eqn.~\ref{eq4} at a range of detunings. Fig.~\ref{NoiseFit} illustrates an example of a frequency noise spectrum filtered for the detection of a spurious RF signal. The minimum axion-photon coupling able to be statistically excluded by the data depends upon the distribution of this residual noise. The treatment of axion field strength follows Daw's convention \cite{Daw1998} and the frequency noise is related to $g_{a\gamma\gamma}$ as per Eqn.~\ref{SySAa}.

Experimental exclusion limits (Fig.~\ref{ExclusionPlot}) were produced via Monte Carlo simulations which were modelled upon the noise statistics of the experimental data. A minimum value for excludable axion-photon coupling was determined from these simulations by injecting axion signals (narrow-band frequency noise) of incremental $g_{a\gamma\gamma}$ (via Eqn.~\ref{SySAa}) upon simulated background spectra and identifying the value at which the threshold was correctly triggered in 95\% of cases. The injected signal assumed a thermalized dark matter halo with a Maxwell-Boltzmann velocity distribution near the Earth with $v_c = 225 \text{ km s}^{-1}$  \cite{Lentz2017}. Further effects on the axion lineshape were built into the simulation based on the drift of the microwave frequencies, which was $\sim$5 Hz per measurement, varying per measurement.
 
To improve SNR, a multi-bin search (similar to Daw's method \cite{Daw1998}) was used, where $n$ translated arrays of $n$-bin averages were examined for axion signals, mitigating the effect of adjacent bin power loss. After binning, data-points above our candidate threshold were isolated for examination, as, according to our noise model, the probability of the background noise breaching this level was 0.05$\%$ (detailed in the supplemental material). A candidate breaching this threshold must be eliminated by examining data taken at an offset microwave detuning.

Data was taken with the TE mode tuned 2.8, 3.1, 3.3, 3.5 and 3.7 MHz above the TM mode frequency, accessing axion frequencies in the MHz range by upconversion. An approximate tuning interval of 300 kHz was chosen to enable tracking of candidate signals between measurements. The experimental data have excluded axions in the probed mass range with an axion-photon coupling exceeding the limits illustrated in Fig.~\ref{ExclusionPlot}.

The next iteration of UPLOAD-CMC will benefit from measurement automation and statistically optimal overlapped binning, width-scaling with axion frequency \cite{Daw1998}. Measurements will begin at the lowest feasible beat frequency supported by the current experiment (about 300 kHz, limited by parasitic coupling between the modes) and progress upwards in frequency space.  Integration time will be increased to days per MHz and the noise filter will be re-calibrated every five tuning steps. Fig.~\ref{ExclusionPlot} illustrates the sensitivities expected to be achievable using frequency-stabilized (as opposed to free-running) loop oscillators at room temperature, as well as a Nb resonator operating at the noise floor of the frequency discriminator at cryogenic temperatures near 4 K, assuming loop-sustaining amplifiers of an effective noise temperature of 8 K \cite{Ivanov2020}, at 30 days per MHz. The projected exclusion limits in Fig.~\ref{ExclusionPlot} were produced by setting the theoretical SNR of various setups to unity and solving for $g_{a\gamma\gamma}$ (detailed in the supplemental material).

As oscillator phase or frequency noise is directly related to axion-sensitivity, increasing phase stability is our most direct avenue of improvement, and is highly possible with modern equipment and techniques, which is discussed comprehensively in the supplemental material. For example, the single sideband phase noise of the oscillator in this work was measured to be approximately $-60$ dBc/Hz at 1 kHz offset (conforming with Leeson's model for this oscillator with an amplifier of $\text{F}=2.6$ at $\text{P}_{\text{amp}}=-33$ dBm and $\text{T}=300$ K), which is far noisier than the state-of-the-art oscillators using specialized frequency-locking and high-Q cavities. For example,  phase noise of $-160$ dBc/Hz at 1 kHz Fourier frequency and at room temperature has been measured \cite{Ivanov2006}; a 10 order-of-magnitude improvement over our experiment. Recently, a path forward to realizing a phase noise of $-185$ dBc/Hz at offsets above 300 Hz has also been proposed \cite{Ivanov2020}.

Our results demonstrate the feasibility of exploiting frequency metrology to develop highly sensitive axion dark matter detectors, with the potential to search wide regions of unexplored axion mass. Proof of principle was achieved with an integration time on the order of just hours, allowing a coupling strength of $\times 10^{-3}$ 1/GeV to be excluded between  7.44 - 19.38 neV. Appraising spectacular modern advances in low phase noise oscillators, a roadmap has been laid to explore untouched axion space with future experiments at room temperature and at cryogenic temperatures using this novel axion-search technique. \vspace{0pt}


This work was funded by the ARC Centre of Excellence for Engineered Quantum Systems, CE170100009, and the ARC Centre of Excellence for Dark Matter Particle Physics, CE200100008, as well as ARC grant number DP190100071. 

\appendix


\onecolumngrid  
\vspace{2\baselineskip}
\section*{\textit{Corrected:} Supplemental Material to: UPconversion Loop Oscillator Axion Detection experiment: A precision frequency interferometric axion dark matter search with a Cylindrical Microwave Cavity}
\twocolumngrid
\vspace{20\baselineskip}

In this supplemental material, we give further technical experimental details and methods not detailed in the main body of the Letter. Following this, we provide the method of determining the UPLOAD-DOWNLOAD axion signal sensitivity as per Eqn. 5 and 8 in the main body of the Letter. The expected axion signal would appear in the spectral density of phase or frequency noise of the experiment as an addition to the loop oscillator's technical  phase or frequency noise. Our sensitivity to an axion signal is therefore limited by our ability to minimize the technical ``background'' noise contributed by the oscillator, so that a small axion contribution may be distinguished above this noise. In this supplemental material we derive the sensitivity of the axion in terms of the spectral density of fractional frequency fluctuations and then derive the expected signal to noise ratio with respect to the oscillator's technical noise. Furthermore, we explain our noise analysis techniques and related Monte Carlo simulations necessary for generating axion exclusion limits. 

\section{Further Experimental Details and Methods}
\begin{figure*}[t]
\centering
\includegraphics[width=16cm]{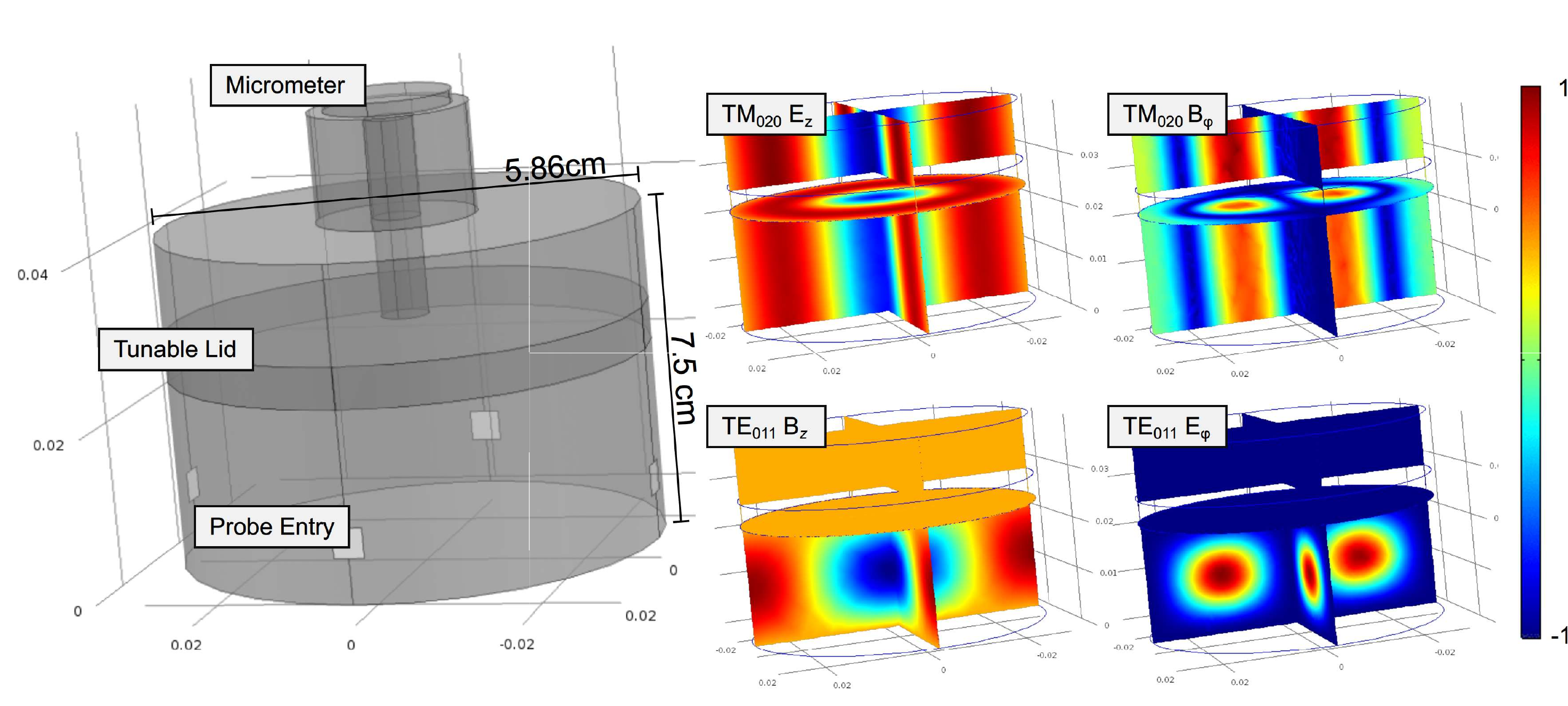}
\caption[comsol]{Schematic of the oxygen-free copper cavity, tunable via a copper lid with displacement controlled by an attached micrometer screw gauge. Numerical solutions (in \textit{COMSOL}) to the $\text{E}_{z}$ and $\text{H}_{\varphi}$ components of the $\text{TM}_{\text{0,2,0}}$ mode and the $\text{H}_{z}$ and $\text{E}_{\varphi}$ components of the $\text{TE}_{\text{0,1,1}}$ mode are shown, whereby the geometric overlap between the E and B fields of the two modes can clearly be seen. In this simulation the TM mode resonates at 8.989 GHz and the TE mode at 9.023 GHz. However, the mode shapes (and consequently, overlap factors) are roughly constant over the experimental tuning range.}
\label{comsol}
\end{figure*}
\begin{figure}[t]
\centering
\includegraphics[width=8cm]{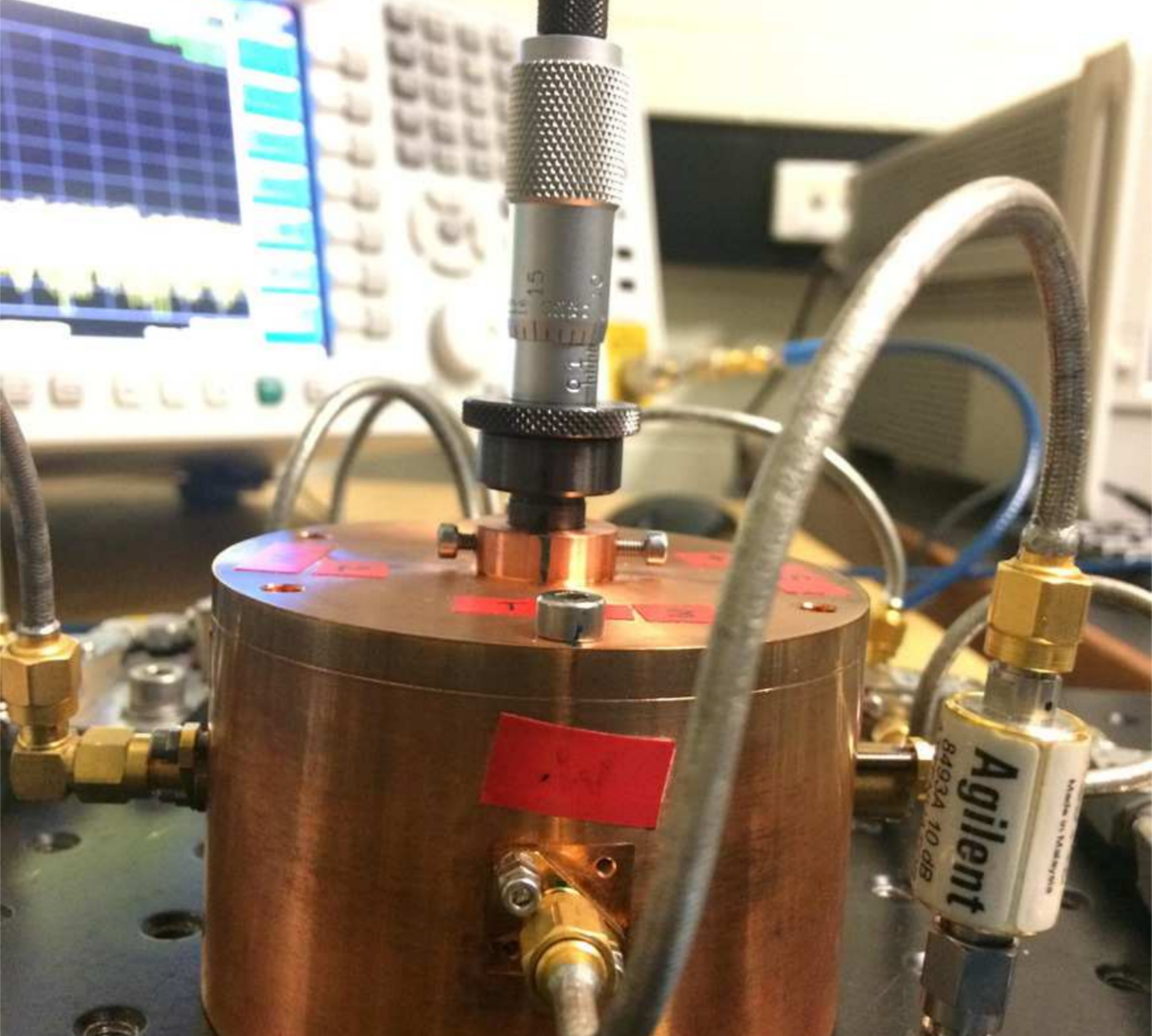}
\caption[UpDownConversion]{The oxygen-free copper cavity supporting $\text{TM}_{0,2,0}$ and $\text{TE}_{0,1,1}$ modes of radius 2.9 cm and height tuning up to 4.5 cm. TE mode tuning is achieved by displacing the lid with a fixed micrometer and modes are excited and probed by four inductive loop probes protruding into the cavity. Unloaded Q is approximately 10,000 at room temperature.}
\label{cavityphoto}
\end{figure}

Figures~\ref{comsol} and~\ref{cavityphoto} and illustrate the oxygen-free copper cavity which supports two electromagnetic modes, via two free-running external loop oscillators, with internal electric and magnetic field structures such that coupling to an axion-like field is expected to occur via the inverse Primakoff effect. The cavity is 2.93 cm in radius and has a functional resonator height tuning from 2.0 - 4.5 cm (aspect ratio tuning from 1.3 - 2.9), achieved by displacing the cavity lid with a screw gauge micrometer. A transverse magnetic (TM) mode in a cylindrical cavity describes a mode which has primarily azimuthally and radially oriented magnetic field components and an axially oriented electric field component. Eqn.~1 of the main paper elucidates the natural selection of a transverse electric (TE) mode to complement the former, in order to maximize axion-photon coupling. A cylindrical TE mode features a primarily azimuthally and radially oriented electric field and only magnetic field in the axial direction. Relevant components of the chosen $\text{TE}_{0,1,1}$ and  $\text{TM}_{0,2,0}$ modes are shown in Fig.~\ref{comsol}, exhibiting significant overlap of the magnetic and electric fields of the respective modes. This overlap enables axion-photon coupling, and is expected to imprint frequency noise on both modes, as detailed in the main paper and in \cite{Goryachev2019}. Other modes exhibiting overlap exist within the cavity, and different resonator geometries and materials are also conceivable to induce the dual-mode axion-photon coupling. These will be the subject of a future paper.

Frequency tuning of the $\text{TE}_{0,1,1}$ mode was achieved by manually altering the height of the cavity lid with a micrometer, while the $\text{TM}_{0,2,0}$ frequency, independent of height, remained practically stationary at 8.999 GHz. Transmission measurements were made by inductively coupling two probes to the $\text{B}_{\varphi}$ component of the TM mode and coupling two probes to the $\text{B}_{z}$ component of the TE mode with coaxial loop probes penetrating the cavity. The beat frequency between the two modes was under constant measurement via an RF mixer connected to a Keysight Frequency Counter (53230A), to quantify the difference and drift in the microwave frequencies (determining targeted axion space).

The frequency noise spectrum of the loop containing the $\text{TM}_{0,2,0}$ mode, which was produced at the output of a frequency discriminator, was interrogated for axion-like signals. While either mode may have been chosen, this mode's stationary resonant frequency reduced the need to tune the frequency noise readout circuit between measurements.

The frequency discriminator comprised a standard phase bridge, which included a phase shifter, a double balanced mixer and another duplicate resonator included as a dispersive element. The reflected and phase shifted signals were set in quadrature (that is, $\varphi_{1}(t)=\varphi_{2}(t)+\frac{\pi}{2}$) such that the output voltage was proportional to the frequency noise of the loop oscillator \cite{Shoaf1973}. For such a system, the oscillator noise can be inferred from the measured spectral density of voltage fluctuations, $S_v(f)$, by
\begin{equation}
S_{y}(f)_{osc}= \frac{S_v(f)}{f_0^2k_f^2G}.
\label{SySv}
\end{equation}
To calibrate the measurement, $k_f$ in volts per Hz must be measured and was determined in this experiment by a known synthesized modulated signal. This slope depends on the $Q$-factor, coupling and incident RF power of the discriminator cavity and can be enhanced with an interferometric scheme, which suppresses the carrier, then amplifies before the RF port. Details of these schemes can be found in the literature \cite{Tobar1994,Ivanov1998,Ivanov2009aa}. As long as the Fourier frequency of interest is smaller than the resonator bandwidth, $k_f$ is effectively a constant value ($k_{f0}$). Otherwise it depends on Fourier frequency as $k_f^2=k_{f0}^2(1+\frac{4f^2}{\Delta f_c^2})^{-1}$. Here $\Delta f_c$ is the bandwidth of the resonant cavity frequency discriminator (3.6 MHz in this experiment) and $G$ is the gain of a low-noise JFET amplifier used after the mixer.

The spectral density of voltage noise ($S_{v}(f)$) measurements from the output of the frequency discriminator were taken at five cavity heights, corresponding to five spans in axion frequency space (2.800 MHz, 3.121 MHz, 3.303 MHz, 3.536 MHz and 3.686 MHz). Ten three-minute traces were acquired at each height, each spanning 10 Hz to 1 MHz in Fourier space, with different resolution in each decade. These spectral densities were measured with a HP89410A Vector Signal Analyzer following amplification by a low-noise JFET amplifier.

\section{Characterising Axion Induced Signal to Noise Ratio}

Standard methods to characterize the frequency noise in oscillators are well known and include both frequency and time domain techniques \cite{NIST1337,Rutman1991,Halford1973}. To measure fast fluctuations quicker than a second we typically analyze the noise in the frequency domain through the spectral density of phase fluctuations. Typically phase fluctuations of an oscillator can be characterized by $S_{\phi}(f)[\text{rads}^2/\text{Hz}]$, the single-sided (i.e. double sideband) spectral density of phase fluctuations or $\mathcal{L}(f)[\text{rads}^2/\text{Hz}]$ the double-sided (i.e. single sideband) spectral density of phase fluctuations, where $S_{\phi}(f)=2\mathcal{L}(f)$. Here, $\mathcal{L}(f)$ is the phase noise with respect to one sideband from the carrier frequency whereas $S_{\phi}(f)$ contains the noise of both sidebands, which collapse and are measured together upon demodulating the carrier to zero. This explains the factor of $2$ difference in value. Accordingly, when speaking of the phase noise associated with an oscillator or component, one must be specific.

When measuring a signal at $f=\pm f_{\text{offset}}$ from the carrier frequency, when we demodulate the signal, it will compete against sideband noise at both $\pm$ offset frequencies and thus against $S_{\phi}(f)$ and not $\mathcal{L}(f)$. Due to the fact that it is impossible to determine if a measured signal is at $f=+f_{\text{offset}}$ or $f=-f_{\text{offset}}$ from the carrier, we must vary the carrier between measurements to observe axion signal translation in the Fourier spectrum. This expected signal translation also allows us to discriminate axion signals from normal RF interference, which in most cases remains constant in Fourier space after retuning. In the event of non-detection of a candidate signal, we can generate exclusion limits at both $\pm f$.

The spectral density of phase noise fluctuations is related to the spectral density of frequency fluctuations by $S_{f}(f)=f^2S_{\phi}(f)[\text{Hz}^2/\text{Hz}]$. However, for frequency noise, it is perhaps more common to represent the spectral density with respect to fractional frequency fluctuations, $S_{y}(f)[1/\text{Hz}]$, which relates to the phase noise via
\begin{equation}
    S_y(f) =\left(\frac{f}{f_0}\right)^2S_{\phi}(f),
    \label{SySp}
\end{equation}
where $f_0$ is the oscillator carrier frequency and $f$ is the Fourier frequency. The axion presents as a narrowband noise source oscillating at the axion frequency, $f_a$, viralized as a Maxwell-Boltzmann distribution of about a part in $10^6$, equal to a narrowband noise source with a frequency spread of $10^{-6}f_a~\text{Hz}$. In this work we represent the spectral density of this axion field by $S_A(f)[\text{kg/s/Hz}]$. To represent the signal in terms of fractional frequency fluctuations, rather than phase fluctuations, one can use the result derived in \cite{Goryachev2019} and combine it with Eqn.~\ref{SySp}, to obtain
\begin{equation}
    \begin{split}
    S_{Ay_{2}}(f)_{\pm} = g_{a\gamma\gamma}^2k_{a\pm}^2S_A(f), \ \
    k_{a\pm}^2= \pi^2\frac{f_1P_{c1}}{f_2P_{c2}}\xi_{\pm}^2, 
  \end{split}
    \label{SySA}
\end{equation}
where $k_{a\pm}$ is the conversion ratio from axion-photon theta angle, $\theta = g_{a\gamma\gamma}a$, to fractional frequency shift, $\frac{\delta f_{a2}}{f_2}$, for axion downconversion (subscript $+$) and upconversion (subscript $-$) respectively. Here, subscript $2$ refers to the readout oscillator, as shown in Fig. 1 in the main text, while subscript $1$ is the pump oscillator, which supplies the second photonic degree of freedom to convert the axion into a signal in the spectrum of oscillator $2$. Furthermore, $P_{c1}$ and $P_{c2}$, are the resonator mode circulating powers driven by the oscillators, while $f_1$ and $f_2$ are the oscillator frequencies. Also, the axion frequency, $f_a$, is related to the Fourier frequency, $f$, via
\begin{equation}
    f_{a_{+}}=f_1+f_2\pm f \ \ \text{or}  \  \   f_{a_{-}}=\lvert f_1-f_2\rvert \pm f,
    \label{Fourier}
\end{equation}
for the axion downconversion and axion upconversion cases respectively.

If we integrate Eqn.~\ref{SySA} on both sides, over the axion bandwidth, we may write the axion-induced signal in terms of induced RMS fractional frequency deviation as
\begin{equation}
    \left<\frac{\delta f_{a_2}}{f_2}\right>_{\pm} = \lvert k_{a\pm}\rvert g_{a\gamma\gamma}\left<a_0\right>,
    \label{RMSyA}
\end{equation}
where the axion field is approximated by $a=a_0\sin(2\pi f_a t)$. The dark matter density, $\rho_{DM}$, has been shown to be related by $\left<a_0\right>=\frac{\sqrt{\rho_{DM}c^3}}{2\pi f_a}$ \cite{Daw1998}, where the cold dark matter density is taken to be $\rho_{DM}=8 \times 10^{-22}\,\text{kg}/\text{m}^3$ (i.e. $0.45\, \text{GeV}/\text{cm}^3$) in this analysis \cite{Daw1998}. Hence, from the measured oscillator noise, $S_{y_{2}}(f)$, the signal to noise ratio (SNR) may be calculated to be
\begin{equation}
    SNR_{\pm} =g_{a\gamma\gamma}\lvert k_{a\pm}\rvert \frac{\left(\frac{10^6t}{ f_{a_{\pm}}}\right)^{\frac{1}{4}}\sqrt{\rho_{DM}c^3}}{2 \pi f_{a_\pm}\sqrt{S_{y_{2}}}},
    \label{eq:SNR}
\end{equation}
assuming the measurement time, $t$ is greater than the axion coherence time so that $t>\frac{10^6}{ f_{a_{\pm}}}$. For measurement times of $t<\frac{10^6}{ f_{a_{\pm}}}$ we substitute $\left(\frac{10^6t}{ f_{a_{\pm}}}\right)^{\frac{1}{4}} \rightarrow t^{\frac{1}{2}}$. Note, for the upconversion case, $f_{a-}$ may be significantly smaller than the microwave carrier frequencies, in which case the SNR is enhanced, being inversely proportional to $f_a$, and the time of integration may be longer before the bandwidth of the axion starts to reduce the sensitivity of the experiment. The opposite is true for the downconversion case, where $f_{a+}$ is much greater than the carrier frequencies. For the experiment at hand, for reasonable aspect ratios of the resonant cavity cylinder, the upconversion case is significantly suppressed (detailed in the next section), while the downconversion case is enhanced, gearing our search toward high mass axions near the sum frequency of the photonic modes. This is not generally true for other modes and geometries, but is beyond the scope of this work.

\subsection{Calculation of Mode Overlap Coefficients}

The axion to fractional frequency fluctuations conversion efficiency, $k_{a\pm}$, is key to our axion sensitivity. This parameter is dependent on the mode overlap coefficients, $\xi_{\pm}=-(\xi_{21}\pm \xi_{12})$, where coefficients $\xi_{12}$ and $\xi_{21}$ are dimensionless, representing the normalized overlap between the two modes as detailed in \cite{Goryachev2019} and defined by
\begin{equation}
        \begin{split}
        \xi_{12}=\frac{1}{\sqrt{V_{1} V_{2}}} \int_{V} \left(\mathbf{e}_{1} \cdot \mathbf{b}_{2}\right) d^{3}r \\
	\xi_{21}=\frac{1}{\sqrt{V_{1} V_{2}}} \int_{V} \left(\mathbf{e}_{2} \cdot \mathbf{b}_{1}\right) d^{3}r.\\
	\end{split}
	\label{overlap}
\end{equation}
Here $\mathbf{e}_{1} $ and $\mathbf{e}_{2} $ are unit vectors representing electric field polarizations of the respective modes and $\mathbf{b}_{1}$ and $\mathbf{b}_{2}$ are unit vectors representing their magnetic field polarizations. The volumes spanned by the modes are $V_{1}$ and $V_{2}$, which may be different in general, but in the case of two modes occupying the same resonator, as in our case, they are the same.

In this work we excite both the $\text{TM}_{\text{0,2,0}}$ mode and $\text{TE}_{\text{0,1,1}}$ mode in a cylindrical cavity, where the height can be modified with a micrometer as shown in Fig.~\ref{cavityphoto}. The $\text{TE}_{\text{0,1,1}}$ mode is the pump mode (mode 1) as shown in Fig. 1 of the main text. A schematic of the cavity is shown in Fig.~\ref{comsol}, along with the overlapping fields. Solutions for electromagnetic modes in a cylindrical cavity maybe derived from Maxwell's equations with the appropriate boundary conditions \cite{Checchin2016}. For a $\text{TM}_{\text{0,2,0}}$ mode in a cylinder of radius $a$ and length $L$, the field vector phasor components are given by
\begin{equation}
        \begin{split}
E_z= E_0e^{i\theta_2}J_0\left(\frac{\chi_{02}}{a}r\right)  \\ 
cB_{\varphi}=i E_0e^{i\theta_2}J_1\left(\frac{\chi_{02}}{a}r\right),
	\end{split}
	\label{TM0,2,0}
\end{equation}
where $J_{i}(j)$ is the $i^{th}$ Bessel function, $\chi_{a,b}$ represents the $b^{th}$ root of the $a^{th}$ Bessel function, $E_0$ is the amplitude of the electric field and $\theta_2$ is an arbitrary phase of the mode. For a $\text{TE}_{\text{0,1,1}}$ mode the vector phasor components are given by
\begin{equation}
        \begin{split}
cB_{z}=cB_0e^{i\theta_1}J_0\left(\frac{\chi_{01}'}{a}r\right)\sin \left(\frac{\pi  z}{L}\right)  \\
cB_{r}=-cB_0e^{i\theta_1}\frac{a\pi}{L\chi_{01}'}J_1\left(\frac{\chi_{01}'}{a}r\right)\cos \left(\frac{\pi  z}{L}\right) \\
E_{\varphi}=-i cB_0e^{i\theta_1}\sqrt{1+\left(\frac{a\pi}{L\chi_{01}'}\right)^2}J_1\left(\frac{\chi_{01}'}{a}r\right)\sin \left(\frac{\pi  z}{L}\right).
	\end{split}
	\label{TE0,1,1}
\end{equation}  
Here $\chi_{a,b}'$ represents the $b^{th}$ root of the derivative of the $a^{th}$ Bessel function, $B_0$ is the amplitude of the magnetic field and $\theta_1$ is an arbitrary phase of the mode.

\begin{figure}[t]
\includegraphics[width=1\columnwidth]{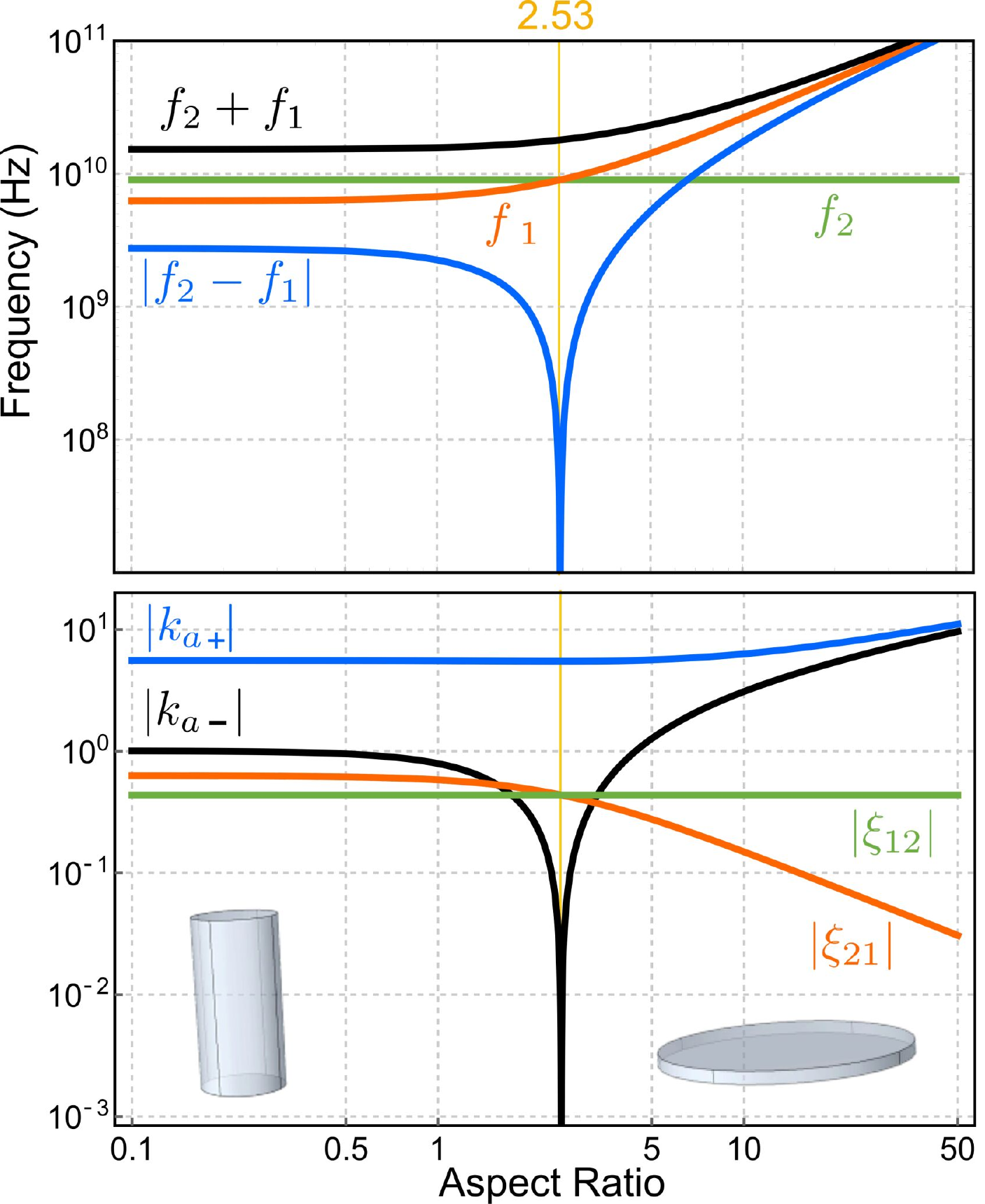}
\caption{{\bf Top:} Frequency versus aspect ratio ($AR$) for the $\text{TE}_{\text{0,1,1}}$ mode ($f_1$) and the $\text{TM}_{\text{0,2,0}}$ mode ($f_2$) when the radius of the cavity resonator is $a=2.93$ cm. The sum frequency, $f_2+f_1$, and the difference frequency, $\lvert f_2-f_1\rvert$, are also plotted, which dictate the axion masses accessible by the experiment. {\bf Bottom:} Mode overlap coefficients, $\xi_{12}$ and $\xi_{21}$, as given by Eqn.~\ref{overlap}, along with the conversion sensitivities, $k_{a\pm}$ for a ratio of circulating power of $P_{c1}/P_{c2}=4$, and as a function of $AR$. In this work the cavity operates with a ratio around $AR$=2.53 as shown by the vertical gold line. Here, the upconversion sensitivity, $k_{a-}$ is suppressed, while the downconversion sensitivity,  $\lvert k_{a+}\rvert=5.5$ remains high, as given by Eqn.~\ref{sens2}.}
\label{Sens}
\end{figure}

To calculate the unit vectors used in Eqn.~\ref{overlap}, Eqns.~\ref{TM0,2,0} and \ref{TE0,1,1} must be normalized to real-valued unit vectors. After the usual amplitude normalization procedure they must be normalized by either $e^{i\theta_{1,2}}$ or $ie^{i\theta_{1,2}}$ to conform to their physical definition in \cite{Goryachev2019}. Using this procedure, we calculate the applicable form of the normalized unit vectors for the $\text{TE}_{\text{0,1,1}}$ mode to be

\begin{equation}
\mathbf{e}_{1}=-\frac{\sqrt{2}J_1\left(\frac{\chi_{01}'}{a}r\right) \sin \left(\frac{\pi  z}{L}\right) }{J_0(\chi_{01}')} \ \hat{\varphi}, \\
\end{equation}
\begin{equation}
        \begin{split}
\mathbf{b}_{1}=\frac{\sqrt{2}\frac{a\pi}{L\chi_{01}'}J_1\left(\frac{ \chi_{01}'}{a}r\right)\cos \left(\frac{\pi  z}{L}\right) }{\sqrt{1+\left(\frac{a\pi}{L\chi_{01}'}\right)^2}J_0(\chi_{01}')} \ \hat{r}  \  \ - \\
\frac{\sqrt{2}J_0\left(\frac{ \chi_{01}'}{a}r\right)\sin \left(\frac{\pi  z}{L}\right) }{\sqrt{1+\left(\frac{a\pi}{L\chi_{01}'}\right)^2}J_0(\chi_{01}')} \ \hat{z},
        	\end{split}
\end{equation} 
and the $\text{TM}_{0,2,0}$ mode to be
\begin{equation}
\mathbf{e}_{2}=\frac{J_0\left(\frac{\chi_{02}}{a}r\right)}{J_1\left(\chi_{02}\right)} \ \hat{z},
\end{equation}
\begin{equation}
\mathbf{b}_{2}= \frac{J_1\left(\frac{\chi_{02}}{a}r\right)}{J_1\left(\chi_{02}\right)} \ \hat{\varphi}.
\end{equation}

From these unit vectors, we may calculate the overlap integrals using Eqn.~\ref{overlap}, for the upconversion, $\xi_{-}$, and downconversion, $\xi_{+}$, cases to be
\begin{equation}
       \begin{split}
\xi_{-}=\xi_{12}-\xi_{21}=\xi_{12}\left(1-\frac{f_2}{f_1}\right),     \\
\xi_{+}=-(\xi_{12}+\xi_{21})=-\xi_{12} \left(1+\frac{f_2}{f_1}\right),  \\
        	\end{split}
	\label{pmoverlap}
\end{equation} 
where
\begin{equation}
\xi_{12}=-\frac{4 \sqrt{2} \chi_{01}'}{\pi  \left(\chi_{02}^2-\chi_{01}'^2\right)},
\end{equation}
\begin{equation}
\frac{f_1}{f_2}=\sqrt{\left(\frac{\pi~AR}{2 \chi_{02}}\right)^2+\left(\frac{\chi_{01}'}{\chi_{02}}\right)^2},
\end{equation}
and $AR$ is the aspect ratio of the cylinder given by
\begin{equation}
AR=\frac{2a}{L}.
\end{equation}
Then substituting Eqn.~\ref{pmoverlap} into Eqn.~\ref{SySA} we obtain
\begin{equation}
k_{a\pm}^2=\frac{32 \chi_{01}'^2}{\left(\chi_{02}^2-\chi_{01}'^2\right)^2}\frac{\beta_{1} P_{1} Q_{L1} (\beta_2+1)^2}{\beta_{2} P_{2} Q_{L2}(\beta_1+1)^2}\frac{(f_2\pm f_1)^2}{f_1f_2},
\label{sens2}
\end{equation} 
where
\begin{equation}
P_{cn}=\frac{4 \beta_{n} P_{n} Q_{Ln}}{(\beta_n+1)^2}.
\end{equation}
Here, $n$ refers to oscillator $1$ or $2$, where $P_{n}$ is the incident power supplied by the oscillator to the resonator input port, $\beta_{n}$ is the input port coupling and $Q_{Ln}$ is the loaded resonator Q-factor.

\subsection{Free Running Loop Oscillator Sensitivity}

\begin{figure}[t]
\includegraphics[width=0.8\columnwidth]{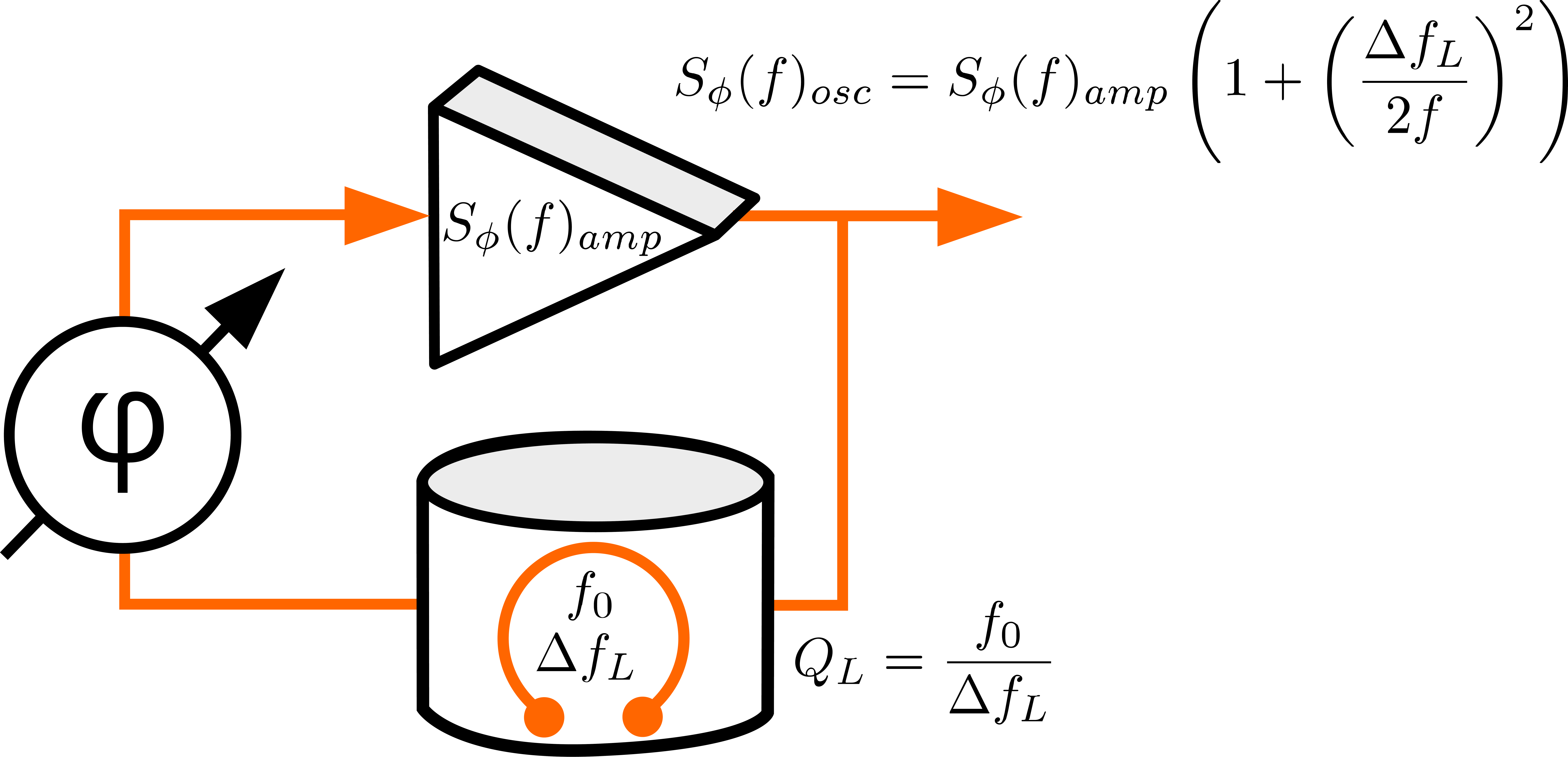}
\caption{Schematic of a simple feedback oscillator, with resonator loaded Q-factor $Q_L$, and amplifier phase noise of $S_{\phi}(f)_{amp}$. Shown is the simple relation to the oscillator phase noise, $S_{\phi}(f)_{osc}$.}
\label{Loop}
\end{figure}

A free-running loop oscillator is the simplest system we can configure and is indeed what we have implemented in this work as our first experiment to to put limits on the axion. In the future, more stable systems can be implemented, as discussed later. The schematic of a simple feedback loop oscillator is shown in Fig.~\ref{Loop}, along with an expression for phase noise. For such a system, the oscillator phase noise is dependent on the phase noise of the amplifier in the sustaining stage, which is typically given by
\begin{equation}
S_{\phi}(f)_{amp}= \left(\frac{Fk_BT_0}{P_{amp}}\right)\left(\frac{f_c}{f}+1\right).
     \label{eq:S1}
\end{equation}
Here, $T_0$ is the ambient temperature, $F$ the amplifier noise figure, $k_B$ Boltzmann's constant, $P_{amp}$ the amplifier input power and $f_c$ the flicker noise corner. To characterize an amplifier in a loop oscillator one has to measure the noise figure and flicker corner as a function of input power \cite{Ivanov2000,Ivanov2014}. At low enough input powers many amplifiers are devoid of a flicker corner. However, in a loop oscillator the amplifier will operate in saturation so it is impossible to operate in this regime using a standard amplifier, and in some cases the flicker noise might have another power law; here it is assumed to be a $1/f$ power law. There are ways to reduce the effect of the flicker noise in an amplifier via the use of clever noise detection and cancellation schemes. This approach was first proposed in 1998 \cite{Ivanov1998} and was coined a ``reduced noise amplifier''. It was only just recently realized in \cite{,Ivanov2020}. This is a possible way to improve these simple loop oscillator systems in the future by many orders of magnitude. In terms of the spectral density of fractional frequency noise, the oscillator noise may be written by substituting Eqn.~\ref{eq:S1} into Eqn. 6 in the main text, also shown in Fig.~\ref{Loop}, and is given by
\begin{equation}
Sy(f)_{osc}= \left(\frac{Fk_BT_0}{4Q_L^2P_{amp}}\right)\left(\frac{f_c}{f}+1\right) \left(\left(2Q_L\frac{f}{f_0}\right)^2+1\right).
     \label{eq:S2}
\end{equation}
Thus, for the free-running experiment, this is the noise we characterize and substitute into Eqn.~\ref{eq:SNR} (with Eqn.~\ref{sens2}) to find theoretical SNR. Note that all experimental data was taken within the corner frequencies ($f_c$) of the loop amplifiers. In this work, we concentrate on upconversion, so that $f_1=f_2+\delta f_{12}$ is tuned slightly away from $f_2$, which remains constant and $\frac{\delta f_{12}}{f_2}<<1$. Also, assuming the pump mode is unity coupled, $\beta_1=1$ (the optimum coupling), to first order in $\frac{\delta f_{12}}{f_2}$ the SNR becomes
\begin{equation}
\begin{split}
SNR_{-}=&g_{a\gamma\gamma}\frac{2.7\left(\frac{10^6t}{f_{a_{-}}}\right)^\frac{1}{4}\sqrt{\rho_{DM}c^3}}{2\pi f_{a_{-}}} \sqrt{\frac{Q_{L2}P_{amp} (\beta_2+1)^2}{(Fk_BT_0)\beta_{2} P_{2}}}\\
&\sqrt{P_{1} Q_{L1}}\sqrt{\frac{1}{\left(2Q_{L2}\frac{f}{f_2}\right)^2+1}}\left(\frac{|\delta f_{12}|}{\sqrt{2}f_2}\right),
\end{split}
\label{eq:FreeSNR}
\end{equation}

\subsection{Stabilized Loop Oscillator Sensitivity}
\begin{figure}[t]
\includegraphics[width=0.8\columnwidth]{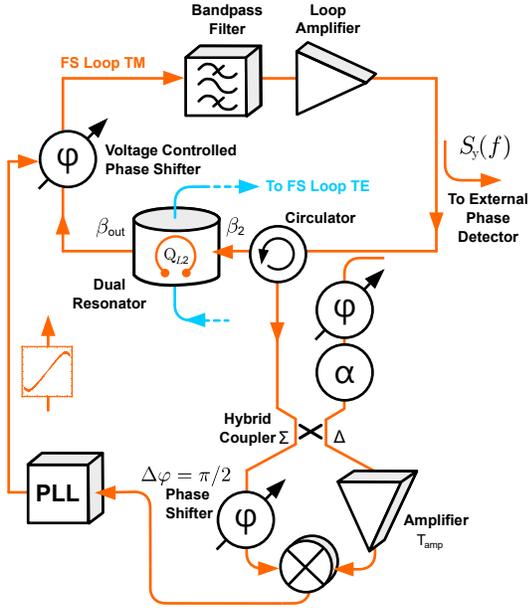}
\caption{Schematic of a frequency stabilized (FS) feedback oscillator with interferometric signal processing (patented). }
\label{FSLoops}
\end{figure}
The lowest noise oscillators are frequency stabilized by a phase detection scheme (such as the system depicted in Fig.~\ref{FSLoops}), which in principle is limited by the effective readout system noise temperature $T_{RS}$ of the internal phase detector (including $T_{amp}$), which is close to ambient temperature for a well-designed system \cite{Ivanov1998}. Given a well-designed system the oscillator noise will be
\begin{equation}
S_y(f)_{osc}=\frac{k_bT_{RS}}{2Q_L^2P_{inc}}\frac{(1+\beta)^2}{4\beta^2} \left(\left(2Q_L\frac{f}{f_0}\right)^2+1\right),
\label{sigma2}
\end{equation}
where $P_{inc}$ is power incident on the input port to the readout mode. Thus, if we use a stabilized loop oscillator, the noise of our system can be significantly reduced, and the SNR becomes (combining Eqns. \ref{sens2}, \ref{sigma2} and \ref{eq:SNR})
\begin{equation}
\begin{split}
SNR_{\pm}=g_{a\gamma\gamma}\frac{16\chi_{01}'}{\left(\chi_{02}^2-\chi_{01}'^2\right)}\frac{\sqrt{\beta_{1}\beta_{2}}}{(\beta_1+1)}\sqrt{\frac{Q_{L1}Q_{L2}P_{1}}{k_bT_{RS}}}\frac{\lvert f_2\pm f_1\rvert}{\sqrt{f_1f_2}}  \\
\times ~ \frac{\left(\frac{10^6t}{f_{a\pm}}\right)^\frac{1}{4}\sqrt{\rho_{DM}c^3}}{2\pi f_{a_{\pm}}\sqrt{\left(2Q_{L2}\frac{f}{f_2}\right)^2+1}}.
\end{split}
\end{equation} 
With the same assumptions as those made in the free-running case, the SNR becomes
\begin{equation}
\begin{split}
&SNR_{-}= g_{a\gamma\gamma}\frac{3.9\left(\frac{10^6t}{f_{a_{-}}}\right)^\frac{1}{4}\sqrt{\rho_{DM}c^3}}{2\pi f_{a_{-}}}\\
&\sqrt{\frac{Q_{L1}P_{1}}{k_bT_{RS}}}\sqrt{\frac{\beta_2Q_{L2}}{\left(2Q_{L2}\frac{f}{f_2}\right)^2+1}}\left(\frac{|\delta f_{12}|}{\sqrt{2}f_2}\right).
\end{split}
\end{equation} 

All projected limits in this work are calculated from the above equation by setting $SNR_-=1$ then calculating $g_{a\gamma\gamma}$. Here $f$ is the Fourier frequency, which we search over, related to the axion frequency by  $f_{a_{-}}=\lvert f_1-f_2\rvert \pm f$, so for every search within the frequency noise spectrum, two axion mass ranges have limits set corresponding to $\pm f$. Note that the readout oscillator (subscript 2) is independent of power in the cavity, as long as the oscillator phase noise is sensitive enough to be thermal noise limited. However, increasing the pump Q-factor and power (subscript 1) will increase the sensitivity. The sensitivity of the readout mode will be increased with Q-factor only if the Fourier frequency of the measurement remains within the bandwidth of the oscillator's resonator, so that $f<\frac{f_2}{2Q_{L2}}$. Thus, to gain bandwidth, the loaded Q-factor may be reduced by over-coupling the readout mode, so that  $\beta_2>1$, where $Q_{L2}=\frac{Q_{02}}{1+\beta_2}$ and $Q_{02}$ is the intrinsic unloaded Q-factor of the readout mode, so $\beta_2Q_{L2}=\frac{\beta_2Q_{02}}{1+\beta_2}$. This assumes extra losses are not present in the over-coupled regime. Here we also ignore the coupling of both oscillator's output port, as they are always kept under-coupled and have little impact on the Q-factor.

Note, this frequency technique is orders of magnitude more sensitive than AC haloscopes which use power techniques \cite{Sikivie2010,Lasenby2020,Lasenby2020b}. This is because precision frequency measurements are inherently more sensitive than power measurements. Moreover, stable cryogenic oscillators have been used successfully for long-term tests of fundamental physics, and have been used to put best limits on variations of fundamental constants and Lorentz invariance violation, with around one year worth of data-taking \cite{Tobar2010,Lo2016,Goryachev2018}, or up to eight years \cite{Tobar2013}. In particular, rotating cryogenic sapphire oscillators have been successfully implemented for long-term measurements, and still maintain the best laboratory recorded limits on Lorentz invariance violations of the photon \cite{Nagel2015}. Here, the rotation effectively chops the signal so it operates at the white fractional frequency noise limit, even in the long term. The success of these experiments gives us confidence to also use cryogenic frequency stabilized oscillators for long-term operation to search for axion dark matter. In our case, the signal is effectively chopped by searching for modulation offsets at Fourier frequencies where the oscillator operates in the white noise limit due to the noise temperature of the frequency stabilization system.

Another advantage of this technique is that when operating in the degenerate mode case (when $f_1\sim f_2$), it is naturally suited to searching for ultra-low-mass axions. We calculate that using the best superconducting resonators \cite{Martinello2018,Buchman1995,Stein1972,Stein75}, sensitivity to the QCD axion is possible for an axion mass range below $10^{-11}$ eV and above $10^{-17}$ eV with just over a year's worth of data. We are pursuing injection locking of the two oscillators to realize this experiment. Recently, another technique suited for low-mass axion searches was published \cite{TOBAR2020}, and could complement this experiment.

To determine the minimum excludable $g_{a\gamma\gamma}$ for a given proposed configuration and measurement time, we follow a similar rationale to Lorentz invariance violation experiments~\cite{Tobar2005}, and to the preceding work~\cite{Goryachev2019}. Specifically, we set the SNR in (Eqn.~\ref{eq:SNR}) to unity and solve for $g_{a\gamma\gamma}$. For an outline of the parameters used for the projections presented in the main paper, see Tab.~\ref{tab:params}.

\begin{table}\centering
\caption{Oscillator Parameters for Projected Limits} 

\begin{tabular*}{\linewidth}{l@{\extracolsep{\fill}}lll}
\toprule
        & \text{FS Cryo Nb}   & \text{FS Copper RT}  \\ \midrule\midrule
$T_{amp}$ 							 &    8 K      &   5 K       \\ 
$T_{amb}$  		&     6 K        &      300 K        \\ 
$P_{inc}$ 		&   100 $\mu$W   &   1 W               \\ 
$Q_{L}$ 		&   $10^9$  &   15000              \\ \toprule \midrule

        & \text{FS Cryo Nb (Degen.)}   & \text{ }  \\ \midrule\midrule
$T_{amp}$ 							 &    8 K      &       \\ 
$T_{amb}$  		&     6 K        &             \\ 
$P_{inc}$ 		&   100 $\mu$W   &             \\ 
$Q_{L_{1}}$ 		&   $10^{10}$  &               \\ 
$Q_{L_{2}}$ 		&   $10^8$  &               \\    
$\beta_{2}$ 		&   $100$  &               \\    \bottomrule

\end{tabular*}
\label{tab:params}
\end{table}

\section{Noise Analysis Techniques and Monte Carlo Simulation}

When producing the exclusion limits in this work, in order to confidently declare non-detection of axions at a certain coupling strength, Monte Carlo simulations were performed, encompassing all known sources of error in the experiment. We will outline the assumptions made in the simulations and the search method employed in the simulations, as upon the experimental data.

The general rationale is as follows: we fit to the experimental frequency noise data and examine the residuals from this fit using a multi-bin search procedure. If a signal with sufficient strength appears in the search spectrum, and survives all cuts and rejections, it is considered as a possible axion signal. Otherwise, simulations are used to determine the weakest axion signal which can be excluded given the noise of the collected dataset. 

The initial fitting to the spectral density of frequency noise data (as depicted in Fig. 2 of the main text) was based upon Leeson's empirical model of oscillator noise \cite{Leeson1966}, and residuals about this fit were examined for excesses of potential axion origin. As the spread of these residuals scales with Fourier frequency (frequencies closer to the carrier naturally experiencing higher absolute phase noise and phase noise spread), the residual spectra were normalized to the Leeson fit, to create the standardized noise spread spectrum depicted in Fig. \ref{fig:gamma} and \ref{binning}. 

The standardized noise from each measurement set (ten traces at one of five detunings), was maximal-ratio combined (based on individual noise spread i.e. SNR) to produce the ``search spectrum'' within which we search for an axion signal. Due to frequency drift (order Hz) in the microwave oscillators, within one measurement set, different traces in fact survey slightly different regions of axion space. Therefore, before averaging, spectra were translated such that they coincided in axion frequency space (rather than Fourier space), using the collected beat frequency data.

\subsection{Monte Carlo Simulations}

\begin{figure}
\centering
\includegraphics[width=8.6cm]{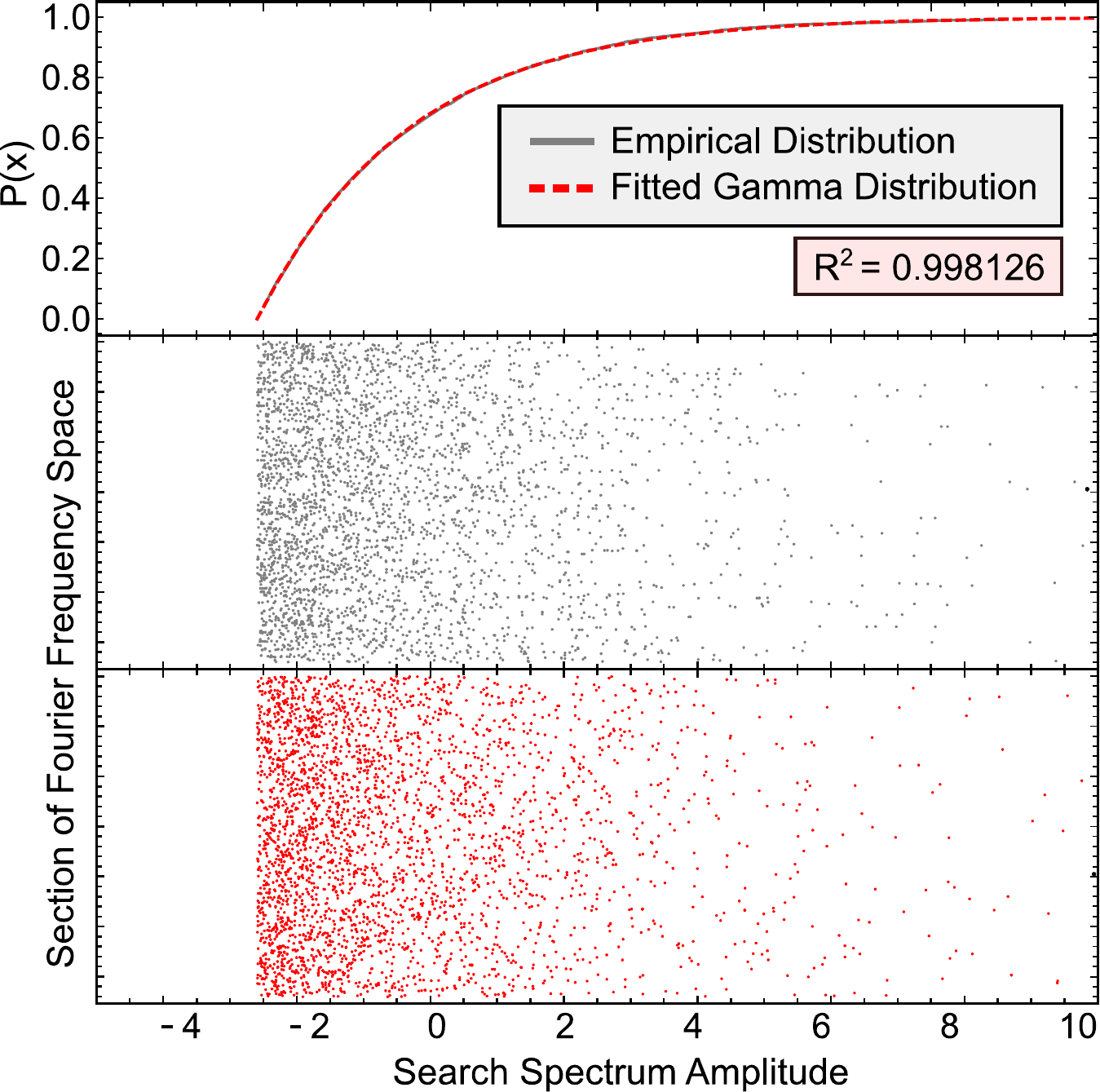}
\caption[UpDownConversion]{\textbf{Top:} An example of fitting a (cumulative distribution function of a) gamma distribution to one span of experimentally acquired spectral density of fractional frequency deviation residuals, which have already been standardized over the relevant Fourier frequency span, such that amplitudes are ``search spectrum amplitudes''. \textbf{Middle:} The experimental data upon which the gamma fit was modelled, from the $f_{\text{TM}} = 8.99888$ GHz and $f_{\text{TE}} = 9.00219$ GHz measurement set. \textbf{Bottom:} A simulated search spectrum produced from the gamma distribution fit, demonstrating high resemblence to parent data. Such simulated spectra, onto which an axion signal is injected are the basis of the Monte Carlo simulations.}
\label{fig:gamma}
\end{figure}
\begin{figure}
\centering
\includegraphics[width=8.6cm]{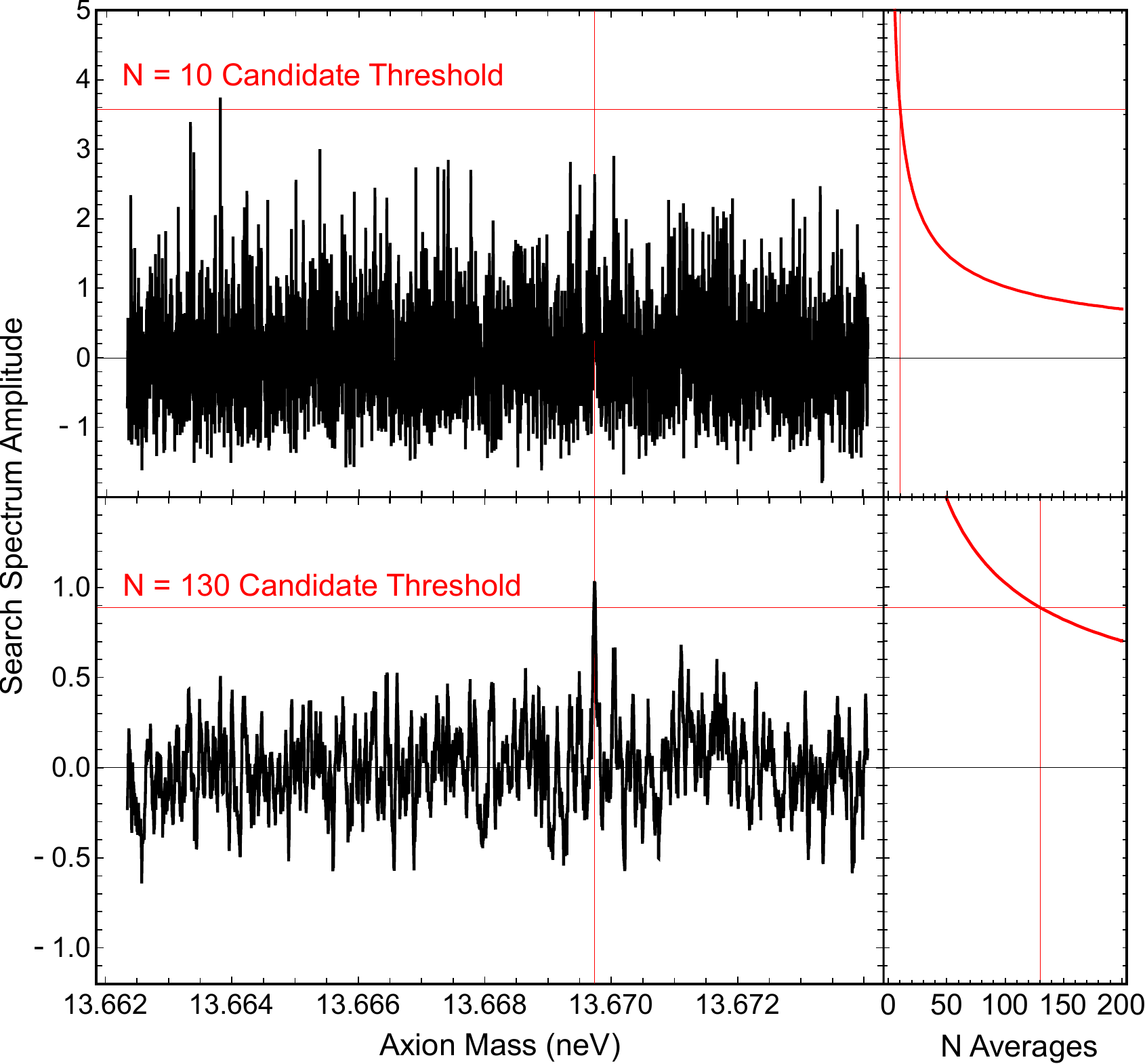}
\caption[Spread]{\textbf{Top-left:} The average of ten simulated search spectra which have been injected with an axion signal of mass 13.67 neV and coupling strength $g_{a\gamma\gamma} = {1.58}\times{10^{-6}}$ 1/GeV. The underlying oscillator noise statistics match the $f_{\text{TM}} = 8.99888$ GHz and $f_{\text{TE}} = 9.00219$ GHz measurement set (as in Fig. \ref{fig:gamma}), and the axion signal is barely visible. \textbf{Top-right:} Determination of the candidate examination threshold for ten averages ($N = 10$), above which 0.05 $\%$ of data is expected to breach. \textbf{Bottom-left:} Result of averaging the 10-averaged spectrum into axion-wide bins, each bin encompassing 13 smaller bins. 13 translated spectra are produced in order to guarantee axion centralization in one bin of the ensemble; all binned spectra are simultaneously represented. Evidently, after averaging into axion-wide bins, the injected axion is discernable above the threshold limit. \textbf{Bottom-right:} Using the threshold function to determine the candidate limit for $N = 130$.}
\label{binning}
\end{figure}

Simulations were performed to determine the necessary strength of ${g}_{a\gamma\gamma}$ that would allow an axion signal to be triggered as a candidate by our search procedure, within simulated datasets resembling our experimental data, with 95\% accuracy (after rejection of false-positives). During data-taking, the Fourier spectra were measured in four spans (decades), each with a different bandwidth, thus each decade has been necessarily independently simulated. Each measurement set (ten traces at one of five detunings) was also independently simulated, as the signal and noise statistics of the constituent traces are unique (although approximately equal).

\begin{figure}
\centering
\includegraphics[width=8.6cm]{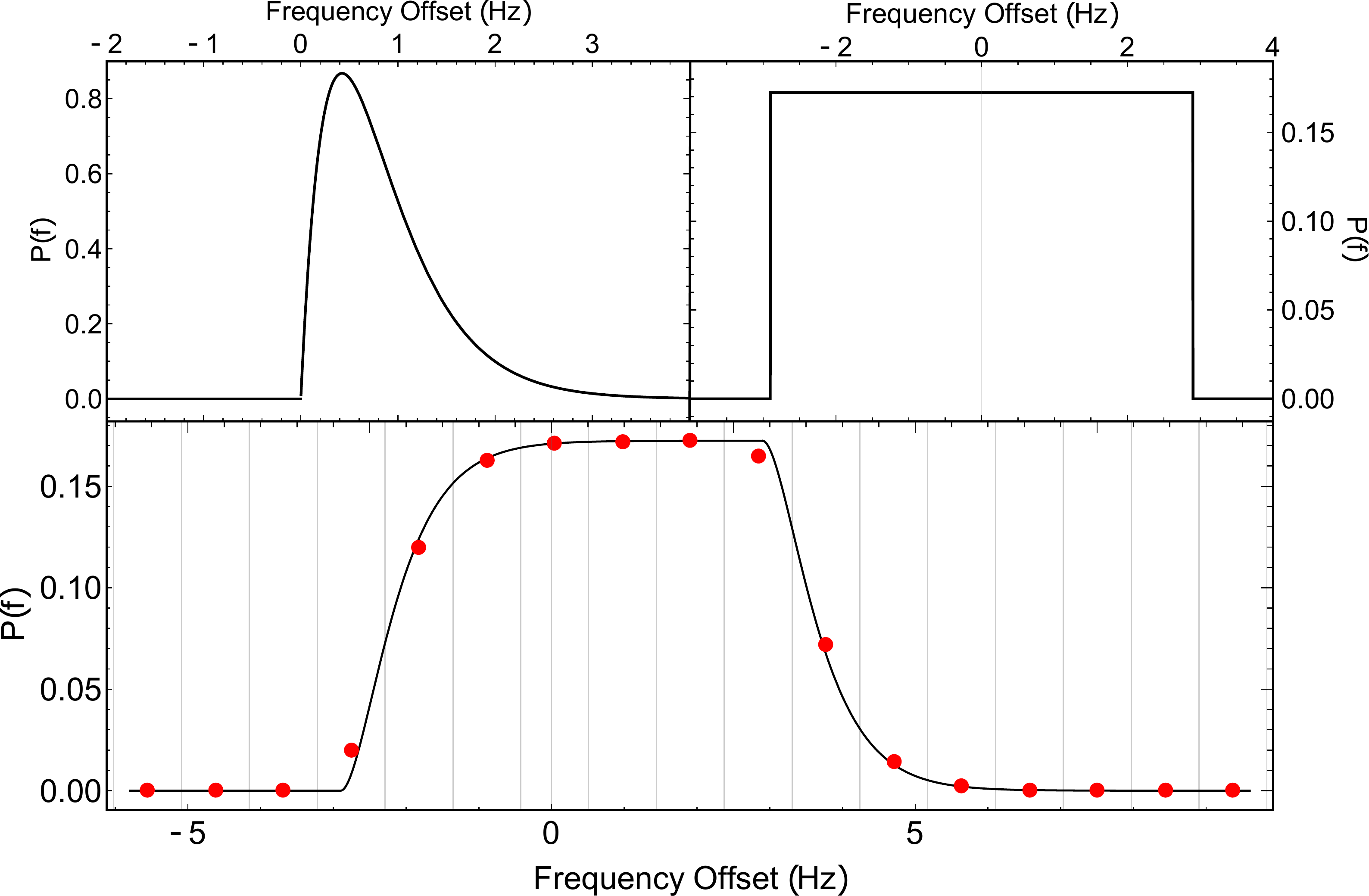}
\caption[Lineshapes]{ \textbf{Top-left:} The axion lineshape for thermalized cold dark matter for an upconverted axion of 7.86 neV (i.e. 1.9 MHz), normalized to unit area. \textbf{Top-right:} Probability density function representation of the uncertainty introduced by drift in the microwave frequencies ($f_\text{TM}$ and $f_\text{TE}$) during acquisition of fast Fourier transforms. In this example, $\delta{f}= \pm 2.9$ Hz. \textbf{Bottom:} The signal lineshape found from convolving the microwave drift distribution with the Maxwell-Boltzmann distribution. The vertical lines represent frequency bin boundaries. As the expected fractional frequency deviation ($\text{Hz}^2/\text{Hz}$) which is induced by the axion field resembles this distribution, in simulations we may integrate within these bins and divide by bin width to determine the spectral density of fractional frequency deviation to be added to the oscillator background before commencing the simulated axion-search. The red dots represent the spectral density value to be added to the oscillator background, where the integral of the entire lineshape represents the total frequency deviation induced by the axion field, proportional to $g_{a\gamma\gamma}^2$, as per Eqn.~\ref{SySA}.}
\label{fig:lineshape}
\end{figure}

The shape of the simulated axion injection assumes a thermalized dark matter halo with a Maxwell-Boltzmann velocity distribution near the Earth with $v_c = 225 \text{ km s}^{-1}$ \cite{Lentz2017}. This distribution would give the interacting axions a kinetic energy kick in frequency space, resulting in a linewidth on the order of $10^{-6}{f}_{a}$. The most significant difference between traces is the associated oscillator drift for each trace, which alters the lineshape of the injected signal (as per Fig.~\ref{fig:lineshape}). During fast Fourier transform acquisition (for a given trace), some thermal drift, spanning $2\delta_{f}$, was observed in the beat frequency between the microwave modes ($f_{\text{TM}}$ and $f_{\text{TE}}$), as recorded by the Keysight 53230A frequency counter, introducing uncertainty in the location of axion frequency space relative to the $f_{\text{TM}}$ mode. To account for this, the Maxwell-Boltzmann axion signal is convolved with a gate function of width $2\delta_{f}$. Due to the variation in $\delta_{f}$ between traces, the measurement sets were independantly simulated, since in cases where drift is higher, the coupled axion power is suppressed, and a marginally higher value of $g_{a\gamma\gamma}$ must be excluded.

A model for the spectral density of technical fractional frequency deviations was produced from the experimental data, as were models for the spread of noise about the fit, for each trace. Empirically, we found that the noise spread statistics resembled a gamma distribution with high conformity (see Fig.~\ref{fig:gamma}).  These models were used to produce simulated spectral densities (in $\text{Hz}^2/\text{Hz}$), ready to be injected with a simulated axion signal, fitted and transformed into search spectra. The simulated densities matched the frequency resolution in each decade of the experimental spectra, which is highly relevant to the expected SNR in each decade for a given axion power input. This explains the nonuniform amplitude of the quoted exclusion limits over all decades of Fourier space (see Fig. 3 in the main work).

The simulated axion fractional frequency noise PSD spectra were combined with the simulated oscillator background spectra by integrating the fraction of the axion signal sitting within each bin-width of the simulated background, dividing by the bin width, and adding the bin-normalized PSD contribution to each point in the oscillator background (see Fig.~\ref{fig:lineshape}) before commencing the fitting and $n$-bin search procedure, as described in the next section. For our data, search spectrum bins ranged from $n$ = 1 to $n$ = 1205 original bins in width, depending on the injected lineshape and the frequency resolution in each decade. The simulation does not assume the optimal positioning of $f_{a}$ with respect to any bin-center.

\subsection{Multi-bin Search Method}
As upon the experimental data, the simulated data is examined for axions using a procedure based upon Daw's six-bin search method as outlined in \cite{Daw1998}. In the simulation, ten axion-injected search spectra are first produced and maximal-ratio combined (to emulate the first step of experimental data analysis). Then, the averaged search spectrum is partitioned into bins of width $n$ roughly matching the linewidth of the expected axion signal (we may use a priori knowledge in the simulation to determine $n$, but the binning procedure was repeated four times on the experimental data to account for the frequency resolution discrepancies between decades). $n$ such $n$-bin search spectra are produced, each time translating the bin center, such that in one spectrum there exists an averaged bin containing the majority of the axion signal, and in one spectrum the signal is $\sim$50$\%$ degraded by loss into an adjacent averaged bin. All $n$ $n$-binned spectra are examined for signals breaching the ``candidate threshold'', which is a function of the number of averages and is unique to each measurement set, having been trained on the noise characteristics of the experimental data. After 10 averages, the percentage for natural breaches of the threshold function stabilizes to 0.05$\%$. A new measurement taken with equivalent sensitivity at a tuning height offset from the $\Delta f$ of the candidate's measurement  may be used to eliminate candidates based on the expected translation of the axion signal in Fourier space. No candidates survived rejections. Theoretically, information from each measurement set may be averaged to produce better limits. However, in this analysis, each set is treated independantly. The superior sensitivity of each set at higher Fourier frequencies suppresses the advantage of set-combining with coincident data taken at a smaller $\Delta f$.

Simulations involved stepping up the axion injection power (the integral of the axion PSD, in $\text{Hz}^2$) until the injected signal was detected above the threshold in at least $95\%$ of cases, after at least 1000 simulations. Exclusion limits on $g_{a\gamma\gamma}$ were produced by equating the successful injection power with the axion component of the total fractional frequency noise spectrum, as per Eqn.~\ref{SySA}. Downconversion limits were excluded from the results paper due to the supressed downconversion sensitivity.

\bibliographystyle{unsrt}

\bibliography{UPLOAD}

\end{document}